\documentclass[10pt,journal,compsoc]{IEEEtran}
\usepackage{xspace}
\usepackage{xcolor}
\usepackage{amsthm}
\usepackage{enumitem}
\usepackage{tabularx}
\usepackage{multirow}
\usepackage{makecell}

\newcommand{\oursystem}{UrbanRama\xspace}

\newcommand{\myparagraph}[1]{\vspace{0.15cm}\noindent\textbf{#1}\xspace}

\ifCLASSOPTIONcompsoc
  \usepackage[nocompress]{cite}
\else
  \usepackage{cite}
\fi
\ifCLASSINFOpdf
   \usepackage[pdftex]{graphicx}
   \graphicspath{{figures/}{pictures/}{images/}{./}}
   \DeclareGraphicsExtensions{.pdf,.jpeg,.png}
\else
\fi
\usepackage{url}

\newtheorem{lemma}{Lemma}

\begin{document}
%
\title{UrbanRama: Navigating Cities in Virtual Reality}
%
%
%
%

\author{Shaoyu~Chen,
        Fabio~Miranda,
        Nivan~Ferreira,
        Marcos~Lage,
        Harish~Doraiswamy,~\IEEEmembership{Member,~IEEE,}
        Corinne~Brenner,
        Connor~Defanti,
        Michael~Koutsoubis,
        Luc~Wilson,
        Ken~Perlin,
        Claudio~Silva,~\IEEEmembership{Fellow,~IEEE}
\IEEEcompsocitemizethanks{\IEEEcompsocthanksitem S. Chen, H. Doraiswamy, C. Brenner, C. Defanti, K. Perlin, C. Silva are with New York University\protect\\
E-mail: \{sc6439, harishd, cjb399, cd1801, kp1, csilva\}@nyu.edu.
\IEEEcompsocthanksitem F. Miranda is with University of Illinois at Chicago.\protect\\E-mail: fabiom@uic.edu.
\IEEEcompsocthanksitem N. Ferreira is with Universidade Federal de Pernambuco. \protect\\
E-mail: nivan@cin.ufpe.br.
\IEEEcompsocthanksitem M. Lage is with Universidade Federal Fluminense. \protect\\
E-mail: mlage@ic.uff.br.
\IEEEcompsocthanksitem M. Koutsoubis, L. Wilson are with Kohn Pedersen Fox Associates PC. E-mail: \{mkoutsoubis, lwilson\}@kpf.com.}
\thanks{Manuscript received October 19, 2020; revised June 16, 2021.}}

\IEEEtitleabstractindextext{%
\begin{abstract}

Exploring large virtual environments, such as cities, is a central task in several domains, such as gaming and urban planning. VR systems can greatly help this task by providing an immersive experience; however, a common issue with viewing and navigating a city in the traditional sense is that users can either obtain a local or a global view, but not both at the same time, requiring them to continuously switch between perspectives, losing context and distracting them from their analysis.
In this paper, our goal is to allow users to navigate to points of interest without changing perspectives. To accomplish this, we design an intuitive navigation interface that takes advantage of the strong sense of spatial presence provided by VR. We supplement this interface with a perspective that warps the environment, called UrbanRama, based on a cylindrical projection, providing a mix of local and global views.
The design of this interface was performed as an iterative process in collaboration with architects and urban planners. We conducted a qualitative and a quantitative pilot user study to evaluate UrbanRama and the results indicate the effectiveness of our system in reducing perspective changes, while ensuring that the warping doesn't affect distance and orientation perception.

\end{abstract}

}

\maketitle

\IEEEdisplaynontitleabstractindextext

%
\IEEEpeerreviewmaketitle

\IEEEraisesectionheading{\section{Introduction}\label{sec:intro}}

\begin{figure*}[t]
	\centering
  	\includegraphics[width=0.9\linewidth]{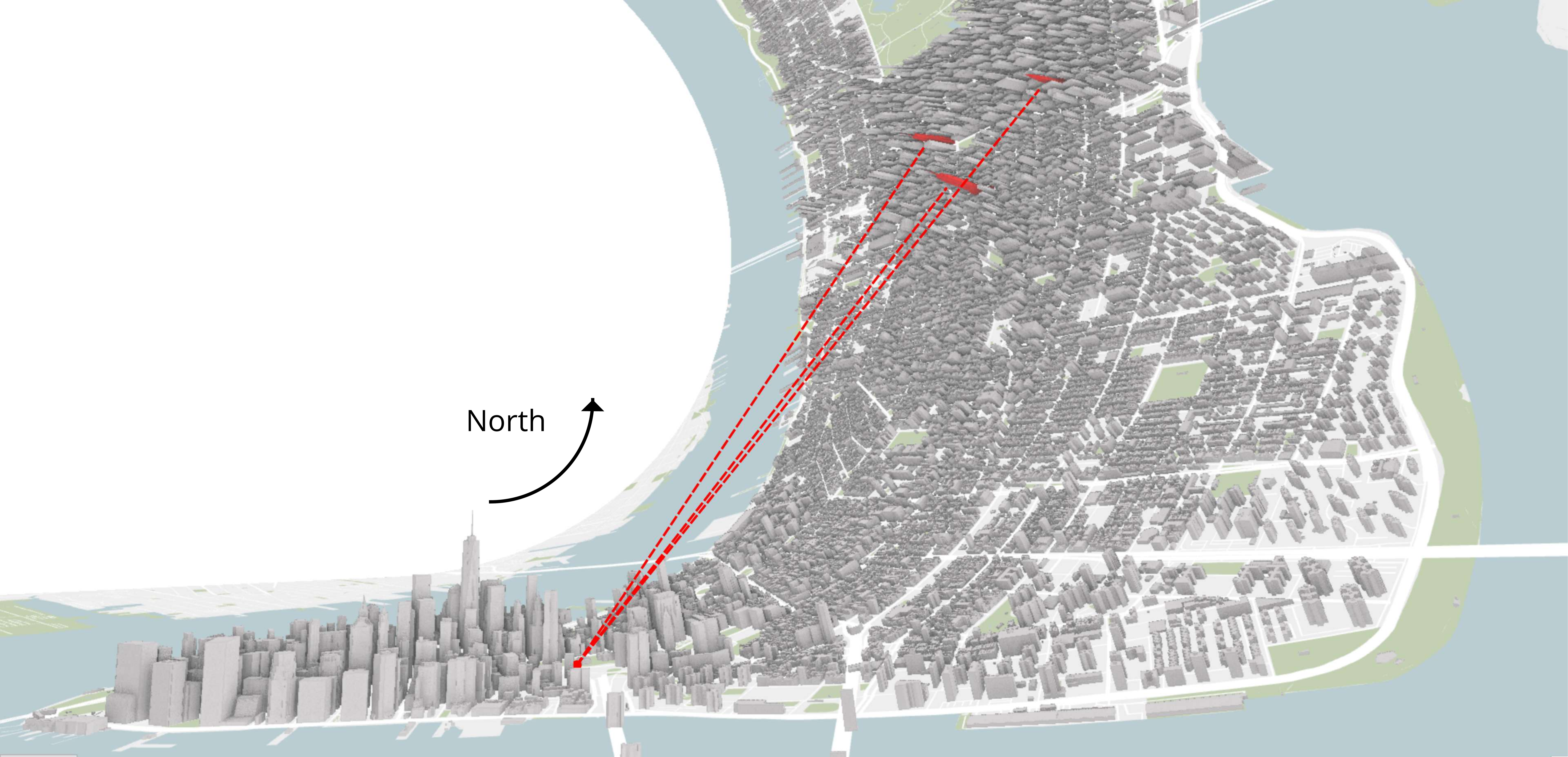}
	\caption{The \oursystem mode in our urban VR system deforms the city allowing the user to obtain both a local as well as a global perspective of the urban environment.
	In this instance, the user is looking north standing in front of New York City Hall in the south of Manhattan, and is able to have a direct line of sight to three important landmarks in Midtown (highlighted): the Bank of America Tower, the Empire State Building, and the Chrysler~Building.}
	\label{fig:teaser}
\end{figure*}

Exploring large virtual 3D urban environments is a central task in gaming and applications related to the domains of architecture and urban planning~\cite{Urbane,Vis-A-Ware,Doraiswamy:2018:IVE:3183713.3193559}. 
The goals in these scenarios are to experience the city and get to know it by \emph{navigating} through the different physical elements present in the virtual world. 
An important consideration in this context is the inherent visual occlusion faced when exploring a city, especially at street level, which restricts the user to a local perspective.
The X-ray or seeing through techniques \cite{Avery2009Xray} may help users mitigate the occlusion problem for short-range local targets.
However, interesting target destinations might happen both at a local as well as a global context~\cite{Urbane}, and such a multiscale perspective of the city is essential for localization, wayfinding, and therefore, for navigation~\cite{christou2016navigation}. 
As a result, users often have to switch between the local~(street-level) and global perspectives~(bird's eye view) during exploration.
This makes the overall process of navigation time consuming and inefficient
\cite{Cockburn:2009:ROZ, Baudisch:2002:KTC}.
Multiple viewports techniques like World-in-Miniature (WIM) \cite{Pausch:1995:NLV} may provide local and global perspectives at the same time.
However, since there is a clear separation between local and global contexts, they are known to have drawbacks, like splitting attention~\cite{Vallance:2001:CPN, Elmqvist:2008:TOM}.
Also, due to the large-scale nature of cities, adding an additional viewport like WIM can raise more issues, such as how to choose its size so that it will not be too small to provide useful information, or too large and affect the major viewport.

The goal of this work is to design a navigation interface for virtual reality (VR) system with the aim of overcoming the above challenges, especially for urban planning and architecture where spatial awareness is a concern.
To do so, as we discuss later in Section~\ref{sec:iterative-design}, we performed an iterative design process alongside collaborators who are professionals in the fields of architecture and urban planning.
Different from ordinary monitor-based systems, VR-based tools 
provide users with a stronger sense of spatial presence, {\it i.e.}, allow users to \emph{experience} the surrounding urban environment as if they were there \cite{hruby2019sound}.
For this reason, VR can greatly help urban design endeavors~\cite{Fisher-Gewirtzman:2018,Sanchez:2017,jamei2017investigating}.
Furthermore, this mode of interaction has been shown to lead to better discovery and collaborative decision making~\cite{Chen:2017:EDS,donalek2014immersive}.
All these features make VR-based visualization systems ideal for urban planning applications with the potential to revolutionize the way development decisions are made in cities.

Based on the interactions with architects and urban planners, we elicited the following list of desired properties for such a system.
\begin{enumerate}
\item[(R1)] Provide a local as well global perspective during the exploration process. This allows users to preserve context without having to constantly switch between different perspectives.
\item[(R2)] Ability to quickly locate and move to building locations / points of interest.
\item[(R3)] Have a small number of controls. This is especially important to not overwhelm users, who, while willing to embrace technology, are not power users.
\end{enumerate}

\myparagraph{Contributions.}
In this paper, we propose a novel deformation approach, called \textit{\oursystem}, that projects the city onto a non-planar view-dependent surface to overcome the location/navigation limitations caused by occlusion. 
In particular, we carry out a user centered deformation of the space, mapping the ground of the city onto the inside of a cylinder.
The approach is inspired by the \emph{Horizonless~Maps} images that were part of the \emph{Here \& There} art exhibit\cite{herethere} and also the ``Rama" starship from Arthur C. Clarke's classic science fiction series.
As illustrated in Fig.~\ref{fig:teaser} and in the accompanying video, such a deformation allows users to get a more detailed global view without affecting the local view.
This method is implemented in a VR infrastructure containing a set of simple and intuitive user navigation controls. 

In order to evaluate our system we performed two user studies (Sections~\ref{sec:user-study} and \ref{sec:new-user-study}) that were approved by the Institutional Review Board~(IRB) of our institution.
The first was a qualitative exploratory study involving professionals from our collaborators' architecture firm to explore the effectiveness of 
this navigation system as well as to gauge the advantages of using the \oursystem deformation during this process.
Participants of this study reported that using \oursystem helps to de-emphasize the need for a high-altitude, bird's eye view in order to navigate. 
In fact, our system allowed them to obtain contextual information 
even from street level, when they assessed the environment from a pedestrian's perspective.
The second study was used to quantitatively assess the effectiveness of \oursystem compared to navigation/localization approaches common in VR systems.
Since exploring large virtual 3D environments is crucial in many domains, different from our first study, we recruited users from a variety of backgrounds to avoid possible bias concerning the professional activities of the participants.
The results from this study demonstrated the effectiveness of our approach in reducing the amount of perspective changes, while at the same time ensuring that the warping does not affect distance and orientation perception.

\section{Related Work}
\label{sec:rel-work}

In this section we review related work from two categories: travel and navigation in virtual reality, and techniques to avoid occlusion in VR.

\myparagraph{Travel in virtual reality.}
According to Bowman et al.~\cite{Bowman:2001:IUI:1246684.1246693}, navigation can be divided into the motor component called travel and the cognitive component called wayfinding. 
There are three main types of travel techniques which do not require physical body movement: teleportation, steering and jumping. In teleportation, viewpoint changes immediately to the destination, while in steering, there is a continuous viewpoint motion from origin to destination. Jumping offers a compromise between teleportation and steering by showing some intermediate viewpoints during the transition.
There are many existing works study the effect of these techniques.
Rahimi Moghadam et al.~\cite{8554159} compared these techniques and found that steering has the best results for spatial awareness but the worst for simulator sickness. However, it was interesting that many users preferred steering regardless of the simulator sickness.
On the other hand, Bozgeyikli et al.~\cite{Bozgeyikli:2016:PTL:2967934.2968105} did not find significant differences between steering, jumping and walking in place in terms of simulator sickness, and Medeiros et al.~\cite{Magic-Carpet} showed that walking in place can induce more simulator sickness comparing with joystick steering.
Weissker~et~al.~\cite{Weissker2018Jumping} showed that active jumping can achieve similar spatial awareness with faster travel time and less simulator sickness, but some users had difficulties to maintain spatial awareness during jumping. They also found that users preferred steering instead of jumping, especially for exploration tasks.
Because of the better spatial awareness and users' preference, we decided to use steering in \oursystem. 
We also decoupled steering into moving within horizontal plane and changing altitude, since \oursystem can provide the local (street-level) and global perspectives (bird's eye view) simultaneously during exploration, which reduces the need for changing altitude.

Several experiments showed that perspective changes can hinder interaction and increase mental burden~\cite{Cockburn:2009:ROZ, Baudisch:2002:KTC}.
Techniques providing both local and global perspectives at the same time are also known as focus+context techniques.
It has been shown that they can improve the comprehension of 3D virtual environments \cite{Pasewaldt:2012:TCD} and avoid splitting attention in perspective changes in 3D virtual environments~\cite{Jankowski:2015:AIE}. 
Furthermore, since there is no separation of focus or perspective change which can lead to confusion (Like user is unsure how the local and global perspectives relate to one another), the continuity between navigation information and direct viewing makes this kind of techniques suitable for tasks where the user's spatial awareness must not be compromised, such as planning and architecture where spatial awareness is a serious concern~\cite{Vallance:2001:CPN}.

\begin{figure}[t]
	\centering
	\includegraphics[width=0.95\linewidth]{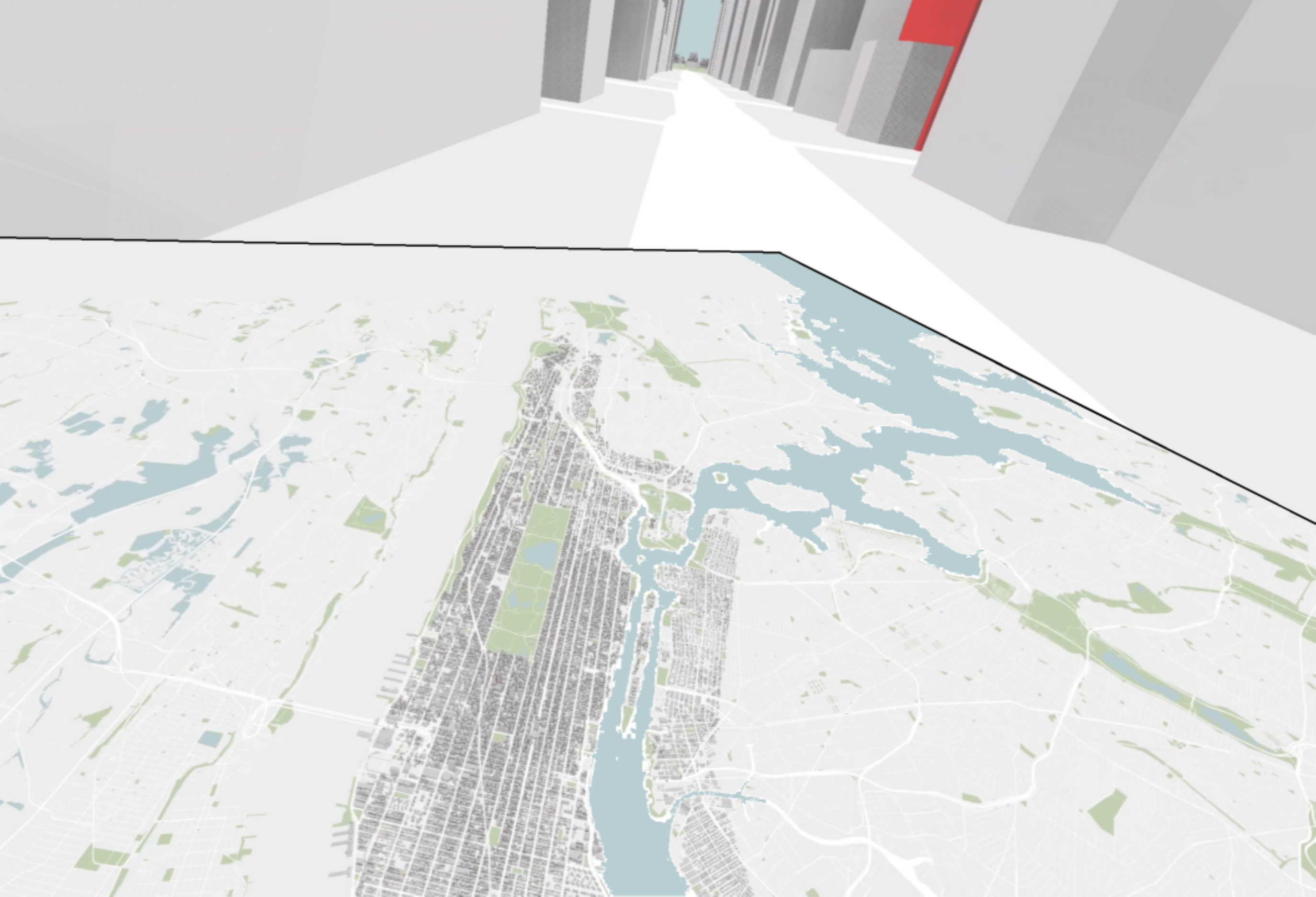}
	\caption{WIM for large scale city. The building selected in the main view and highlighted in red can hardly be found in WIM.}
	\label{fig:WIM}
\end{figure}

There are several navigation techniques which compromise between local and global navigation contexts. 
However, they all have different drawbacks make them not suitable for urban related tasks.
Tan et al.~\cite{Tan:2001:ENC} coupled speed control to height (position) and tilt (viewpoint) control, such that users will see a local view while standing still or moving slowly, and a global view while moving fast, to enable seamless transition between local and global views.
However, users need to move fast to enter the global view, which can be undesired because during exploration and analysis, users usually do not have a specific point of interest in mind and prefer to stay still.
Also, it cannot provide both local and global views at the same time.
Krekhov et al.~\cite{Krekhov:2018:GWT} enabled seamless transition between local and global views by changing users' size in VR.
But it also cannot provide both local and global views at the same time and requires perspective change.
The World-in-Miniature (WIM) metaphor~\cite{Pausch:1995:NLV, LaViola:2001:HMN, Elvezio:2017:TLH}, which shows the global context in a miniature 3D map thus providing the local and global contexts at the same time.
However, there is a clear separation between the local and global contexts. 
This can lead to split attention or confusion on how the local and global perspectives relate to one another~\cite{Vallance:2001:CPN}.
In addition, due to the large scale nature of urban scene, WIM can be too small to provide meaningful context to the users, as shown in Fig.~\ref{fig:WIM}.
WIM is known to have complexity and occlusion management issues \cite{Trueba:2009:COM} and large scale nature of urban scene can make them worse.
WIM has only been tested in small city with very few blocks~\cite{Elvezio:2017:TLH} and it remains an unsolved problem for applying WIM in the large-scale urban scene.
Fukatsu et al.~\cite{Fukatsu:1998:ICB} used bird's eye view instead of WIM for the additional viewport, but it still has similar issues.
Elmqvist and Tsigas~\cite{Elmqvist:2008:TOM} proposed a taxonomy of 3D Occlusion Management for Visualization.
In the taxonomy, the above techniques are categorized into "Multiple Viewports" category, while our proposed method falls into a different "Projection Distorter" category.

\begin{figure*}[th!]
    \centering
    \includegraphics[width=0.45\linewidth]{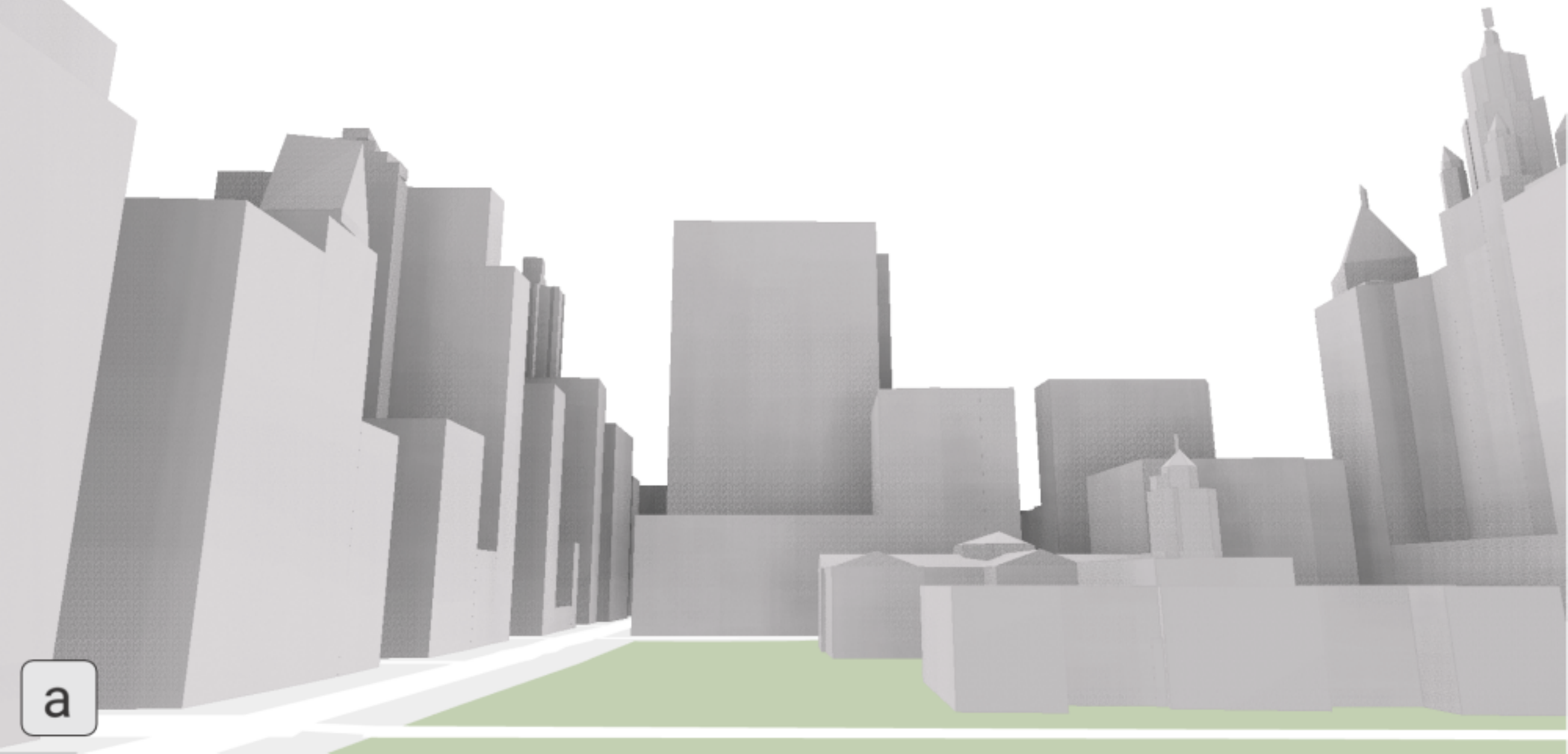}
    \includegraphics[width=0.45\linewidth]{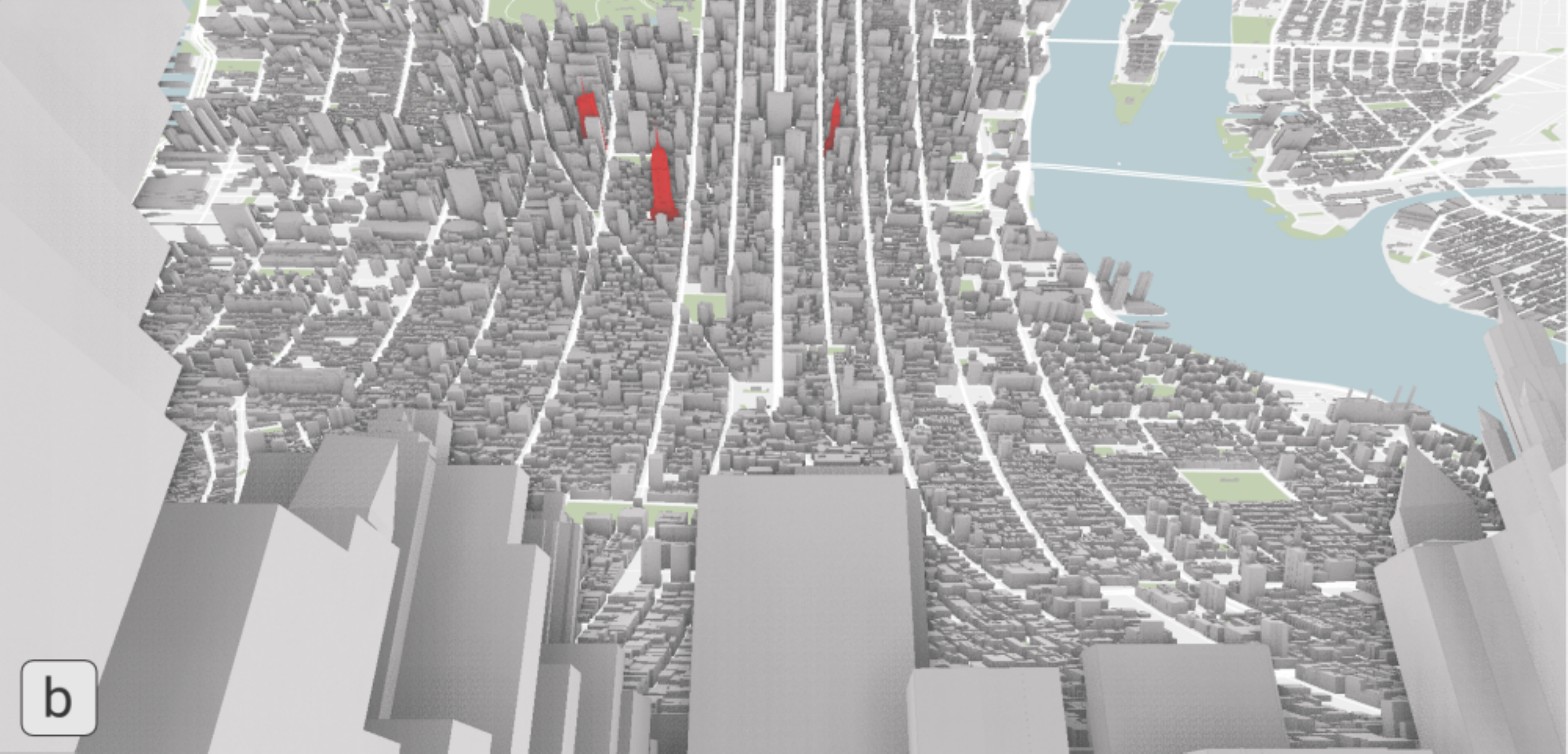}
    \caption{Comparing the normal view a user would have with the view using \oursystem mode when standing in front of City Hall and looking north. 
    (a)~Note that in the normal view, the surrounding buildings occlude all details of other areas beyond this neighborhood. (b)~When in \oursystem mode, users can easily see faraway landmarks such as the Empire State Building, Chrysler Building, and the Bank of America Tower (highlighted). Note that the deformation shown in Fig.~\ref{fig:teaser}
    corresponds to the user being in the position shown in this figure.
    }
    \label{fig:comparison}
\end{figure*}

\myparagraph{Multi-perspective Views in 3D Virtual Environments.}
Multi-perspective projections \cite{Moser:2008,Lorenz:2008,Pasewaldt:2011,Wu:2018} and view dependent deformations~\cite{tong2016view, tong2017glyph} have been used to solve the well-known occlusion problem when visualizing 3D environments.
One advantage of these approaches is that the information bandwidth of user's view is increased by integrating multiple perspectives.
A user study~\cite{IEEEVR-occlusion-study} compared three disocclusion methods in VR (top view, X-ray and multi-perspective) and found that multi-perspective generally performs better than the other two methods.

Some of these approaches provide local and global perspectives at the same time for urban scenario.
Spur and Tourre~\cite{Spur:2018} projected a 2D data layer to a sphere surrounding users to utilize the unused screen space and allow users to have an overview of distant data when the user is at the street level. 
But this method retains the 3D buildings as is without projecting them, due to which the spatial context which is essential for navigation~\cite{christou2016navigation} tasks, is lost.
Pasewaldt~et~al.~\cite{Pasewaldt:2011} proposed a deformation of city and landscape models 
based on b-splines with the goal to get views from multiple perspectives and maximize screen space usage.
This approach was designed to enable a bird's eye view based navigation of the landscape and therefore it is not useful for a street level experience. 
Furthermore, the deformations applied to the building shapes also inhibit navigating to distances too far from the user position.
The \oursystem strategy for deforming and projecting the city's geometry was carefully designed to minimize building shape deformations and also provide both local and global perspectives of the city landscape.

\section{\oursystem Space Deformation}
\label{sec:projection}

Our main idea in trying to ease user navigation around an urban space is to deform the city in order to provide both a local (street-level view) and global perspective (bird's eye view) in the same view.
We accomplish this by deforming the city's geometry using a cylindrical projection based on the user's point of view. 
Fig.~\ref{fig:comparison} shows how 
this method avoids local occlusion and enables users to see far away buildings similar to a bird's eye view.

After trying out the deformation using both the sphere as well as the cylinder as the base of our proposal, we decided to use the cylinder-based deformation. 
The reason for this choice was that in the initial design iteration, our collaborators found that the spherical approach, when deforming all around the user, was disorienting and making it difficult for users to place themselves.
On the other hand, the cylindrical approach maintains the flat context in the neighborhood of the user's position.

Our projection method is based on the stereographic projection\cite{nla.cat-vn2516475}, which is a conformal map projection that maps points on the surface of a sphere to points on a plane.
We first map points from a plane to points on the surface of a cylinder. 
Without loss of generality, let the cylinder lie tangentially on the plane and the intersection line with the plane is the $y$-axis
so that its axis is parallel to the $y$-axis, and has diameter $d$.

Then, the projected coordinate $(x,y,z)$ of a point $(X,Y,0)$ on the plane on this cylinder is computed as:
$$ (x,y,z) = (\frac{d^2 \cdot X}{d^2+X^2},Y,\frac{d \cdot X^2}{d^2+X^2}) $$

\begin{table*}[h!]
\caption{Overview of the three different versions of the system. Version 1 was used in our initial meetings to elicit the system requirements. Version 2 was used in the qualitative user study. Version 3 was used in the quantitative user study.}
\centering
\begin{tabular}{l|l|l|l}
 & Version 1 & Version 2 & Version 3 \\
Controller used for navigation & Both & Right hand only & Right hand only \\
Controller navigation DOF & 6 DOF & 3 DOF (translation only) & 3 DOF (translation only) \\
Separated altitude change & No & Yes & Yes \\
Altitude after flying through & 100m & 100m & not changed \\
Flying time & Constant & Constant & Proportional to distance \\
Cylindrical axis follows user's head & No & Yes & Yes, with threshold and option to pause
\end{tabular}
\label{tab:versions}
\end{table*}

We now describe how to extend the above approach when the scene contains the buildings of the city. Unlike the case of a plane, a point in the scene can have a non-zero height.
To deform such scenes, consider a point $p$ in the scene having coordinates $(X,Y,Z)$. 
We first map the point $p' = (X,Y,0)$  (that lies on the plane) to the cylinder using the above mentioned formula. 
Let this point be $q'$. We then translate $q'$ $Z$ units towards the axis of the cylinder.
\begin{figure}
	\centering
	\includegraphics[width=0.9\linewidth]{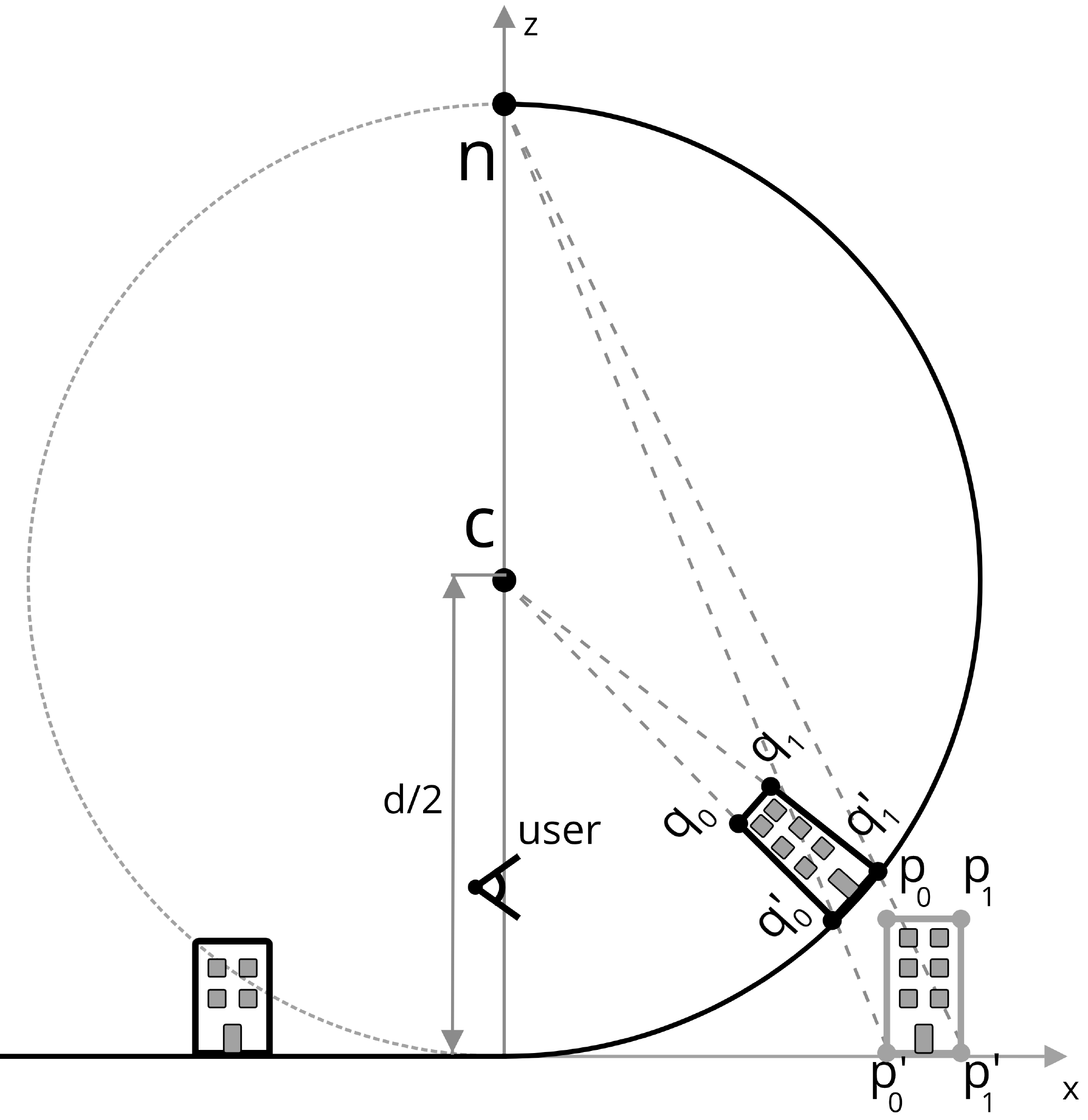}
	\caption{Extending the cylindrical projection approach to deform a city. The key idea is to project the points at ground level onto the cylinder, and deform the elevated points towards the central axis of the cylinder. Note that such a deformation towards the central axis ensures that buildings do not intersect with each other.}
	\label{fig:building}
\end{figure}
For example, to deform a point $p_i = (X,Y,Z)$ using cylindrical approach, first we calculate the projected point $q_i'$ for $p_i' = (X,Y,0)$, which is given by
\begin{equation}
    q_i' = (x,y,z) = (\frac{d^2 \cdot X}{d^2+X^2},Y,\frac{d \cdot X^2}{d^2+X^2}).
    \label{projection}
\end{equation}
Then, we calculate the direction $\overrightarrow{d}$ towards the axis, which is $(-x,0,d/2-z)$ and translate the point $Z$ units along the normalized direction $\overrightarrow{d}$. 
Then we have the projected point $q_i$~(see Fig~\ref{fig:building}).
When using the cylindrical approach, the result looks like bending the city plane as a fabric.

When using the deformation, which we call the \textit{\oursystem mode} or simply the \textit{Rama mode}, the axis of the cylinder is placed directly above the user's position on the street at a height $\frac{d}{2}$,
where $d$ is the diameter of the cylinder. 
The axis is oriented orthogonal to the eye direction and the up vector $(0,0,1)$. 
As shown in Fig.~\ref{fig:building}, we deform only half of the cylinder that is in front of the user. 
This is because camera position has the same $x$ and $y$ coordinates as the cylinder center and the scene behind the viewer is clipped during the rendering. 
Figs.~\ref{fig:teaser} and \ref{fig:comparison}(b) show the deformation of NYC when the user is looking north, standing in front of City Hall.
The following lemma helps us select parameters to ensure that the deformed buildings do not intersect.
\begin{lemma}
Two non-intersecting 3D line segments \textit{will not intersect} after deformation if the maximum $z$ value of the lines is less than $d/2$.
\end{lemma}
\vspace{-0.1in}
\begin{proof}
Consider two points $P_1 = (X_1, Y_1, Z_1)$ and $P_2 = (X_2, Y_2, Z_2)$ such that $P_1 \neq P_2$. Assume that they map to the same point once deformed (i.e., it is an intersection point in the deformation). 
By definition, the deformation does not change the $y$-coordinate. Thus, if the two points intersect after deformation, then $Y_1 = Y_2$.
Without loss of generality, let $Y_1 = Y_2 = 0$, and let the camera (user) be located at the origin. Also, assume that both these points are in front of the camera.
Consider the previously described deformation approach. First the points $(X_1, 0, Z_1)$ and $(X_2, 0, Z_2)$ are projected onto the circle with diameter $d$ centered at $(0,0,\frac{d}{2})$. By definition, this mapping is one-to-one. Thus, the two points are mapped to a unique point on the above circle. Next, the mapped point on the circle is extruded by $Z_1$ and $Z_2$ units respectively, towards the center. 
The vector representing the extrusion for a given point is essentially the radius line (perpendicular to the tangent at the mapped point). Thus, two such lines can intersect only at the center. Thus, for the deformed points to be the same implies that the extrusion, which is equal to $Z_1$ (and $Z_2$) is $\frac{d}{2}$.

Now consider the case when one of the points, say $P_2$ is behind camera. Since there is no deformation for $P_2$, the only possibility for the deformed $P_1$ to be equal to $P_2$ is when this deformed point is behind the camera. In other words, the amount of extrusion $Z_1 > \frac{d}{2}$.
\end{proof}
\vspace{-0.1cm}
Thus, by choosing a suitable cylinder radius, one can ensure that such a deformation will not create self-intersecting geometries. 
In particular, we set $d=5$~kilometers. 
This value was found experimentally to provide a good balance between low local deformation while providing an experience similar to the bird's eye view for faraway regions.
Finally, since this value is much larger than the buildings' heights the shape deformation incurred in the buildings' geometries (as illustrated, for example, by difference in length between $\bar{p_0p_1}$ and $\bar{q_0q_1}$ in Fig.~\ref{fig:building}) is small and visually negligible.

\section{VR Infrastructure and Iterative Design Process}

\begin{table*}
\caption{Overview of controls used in the different versions of the system.}
\centering
\begin{tabular}{l|l|l|l}
 & Version 1 & Version 2 & Version 3 \\
Right thumbstick & Move along controller pointing direction & Change altitude & Change altitude \\
Button A & Enter Rama mode & Enter/exit Rama mode & Enter/exit Rama mode\\
Button B & Exit Rama mode & Move forward & Move forward\\
Right trigger & Fly to selected point & Fly to selected point & Fly to selected point \\
Right grip & N/A & N/A & Pause/resume Rama mode \\
Left thumbstick & Rotate camera & N/A & N/A \\
Button X & Increase Rama mode radius & N/A & N/A \\
Button Y & Decrease Rama mode radius & N/A & N/A 
\end{tabular}
\label{tab:controls}
\end{table*}

\label{sec:iterative-design}

In this section we detail the iterative design process of our prototype urban virtual reality system. 
To better frame the presentation, we first discuss the hardware setup used.

\begin{figure}[b]
	\centering
	\includegraphics[width=0.8\linewidth]{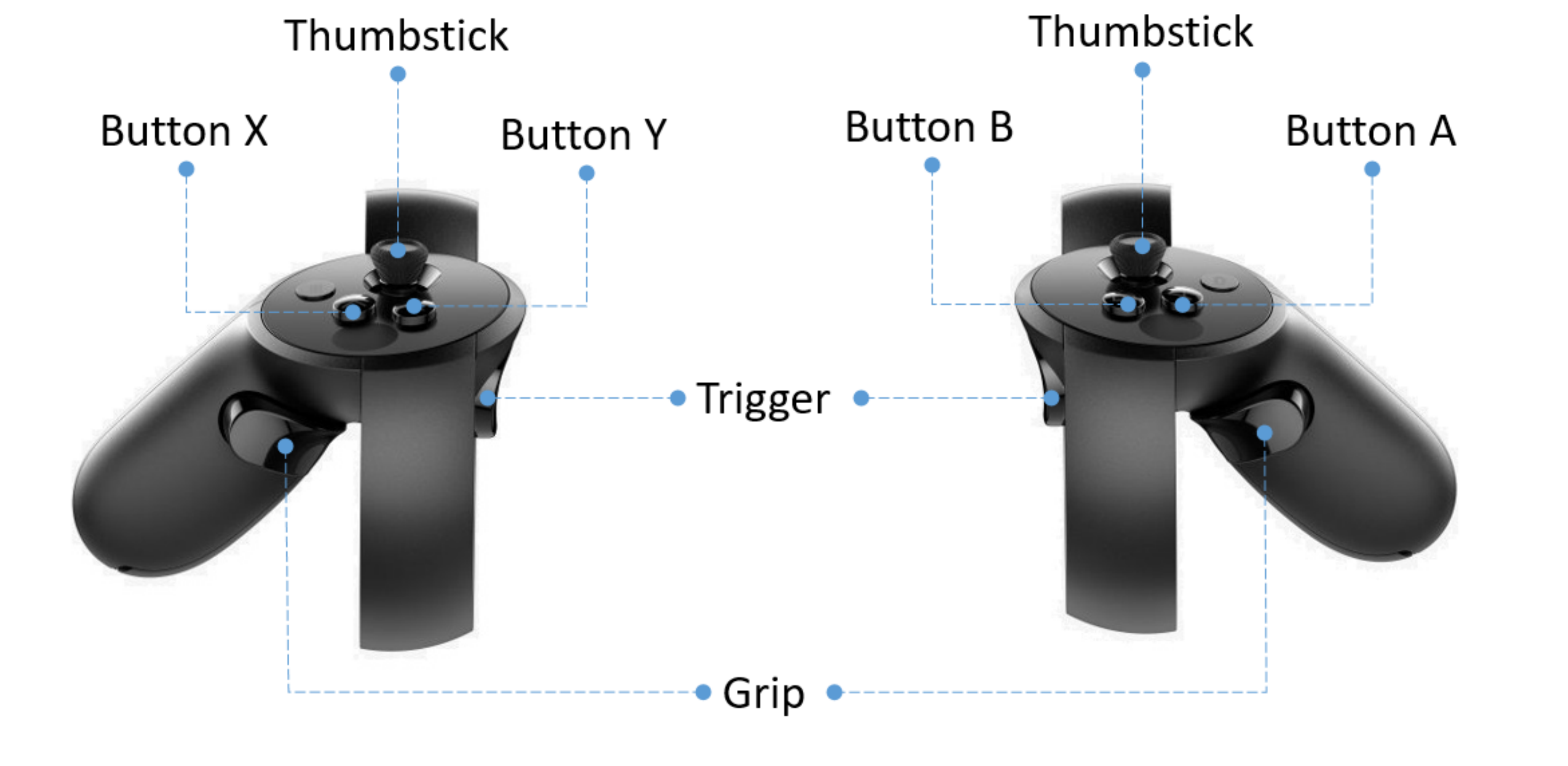}
	\caption{The different controls provided to the user by the Oculus Rift controllers. We primarily use only the right controller in our system.
    }
	\label{fig:controller}
\end{figure}

\subsection{VR Infrastructure}
\label{sec:navigation}

Our system uses the Oculus Rift, a popular VR headset~\cite{Steam}.
It has a $2160\times1200$ resolution ($1080\times1200$ per eye) OLED panel with 110 degree field of view and 90 Hz refresh rate,
and comprises of two controllers (Fig.~\ref{fig:controller}) that are used to interact with the virtual environment, in addition to a headset which tracks the users view. Note that \oursystem only uses the right controller.

\subsection{Design Process}
Our iterative design process consisted of three main phases.
After each phase we developed an improved version of our system based on the input/feedback received in the previous phase.
The main differences among the versions are summarized on Table~\ref{tab:versions} and on Table~\ref{tab:controls}.

The first phase consisted in a series of meetings with our three domain collaborators, during which we elicited the main systems requirements R1, R2 and R3, mentioned in Section~\ref{sec:intro}. 
After that we developed the first version of interface which had a goal to give as much navigation freedom as possible to the user (Version 1 in Table~\ref{tab:versions}).
In particular, this interface made use of both the left and right controllers as follows.
The thumbstick on the right controller was used to move along the direction the controller was pointing to. 
Thus the users could move in arbitrary directions. 
Buttons A and B were used to toggle on and off, respectively, the Rama mode. 
Also, in this version, the cylindrical axis was fixed when the Rama mode was toggled on.
To change the orientation of the cylindrical axis, the user had to turn off Rama mode, orient herself in the required direction, and turn on the Rama mode again.
The two buttons on the left controller allowed the user to increase and decrease respectively, the radius of the cylinder.
Also, the left controller thumbstick allowed the user to rotate the camera.
Finally, in addition to the button controls, the right controller also acts as a laser pointer. 
We use this to allow users to point to locations of interest.
For example, pointing to a particular building selects and highlights it. 
Pointing to a location or building and squeezing the trigger will move the user to the desired target.
The movement happens as a fly through transition.
In case this is a building, then the user is moved to a point near that building.

In the next phase of our design process, we had a new series of meetings, with our domain expert collaborators, during which they got the chance to use our prototype and comment on it.
The main feedback we received concerned the navigation which they found too cumbersome. 
Moreover, the ability to rotate the camera turned out to be unsettling for some of the users.
Based on this feedback, we added another design requirement that tries to simplify and minimize the controls in our system. Also, based on their reported use case scenarios, we decided to restrict the movement freedom provided to the user. In order to simplify the navigation, the feedback from them allowed us to define 5 different altitude levels that were implemented in our next version of \oursystem: 5m (street level), 100m (higher than most normal buildings), 500m (higher than most skyscrapers), 1km and 2km (bird's eye view). We~also set the cylindrical axis in Rama mode to continuously change with the headset movement.
We then performed a few more iterations to improve our use of controllers and interaction parameters, resulting in Version 2 described in Table~\ref{tab:versions}.

Using this version, we moved to the next phase of the design process, during which we conducted a qualitative user study (Section~\ref{sec:user-study}) to collect feedback from a broader group of domain users.
The collected feedback was then used to fine tune the system, resulting in the third version of \oursystem. This version is detailed in Section~\ref{sec:system}, and was the prototype used in a quantitative user study, detailed in Section~\ref{sec:new-user-study}.
Throughout this process, no user participated in more than one phase of our study.

\section{Qualitative User Study}
\label{sec:user-study}

As an initial step to evaluate how experts perceive the usefulness of \oursystem, we conducted an exploratory user study with architects and urban planning professionals. 
Participants were asked to complete a series of tasks designed to mimic potential navigation and localization use cases, and elicit interactions and judgments about the interface. 

\myparagraph{Participants}
The participants in this study were 11 
professionals: architects, urban planners, computational designers, and experts in the use of urban analytics from Kohn Pedersen Fox, a large architecture firm located in NYC.
Their experience in their roles ranged between 6 months and 15 years.
Four participants had only minimal previous experience with VR, while seven participants were experienced users, including one VR developer. 
All participants were familiar with the geography and landmarks of NYC.  
The domain collaborators that participated in the initial phases of our design process were not included in the group of participants in this study.

\subsection{Methods and Data Collection}
To capture users' reasoning during tasks and continuous evaluation of the system, a think-aloud protocol was used. 
A semi-structured questionnaire was used to solicit specific responses to features of the tasks. 
Participants' responses were audio recorded throughout the session.
The study session protocol included the following steps. After being introduced to the virtual reality hardware (Oculus Rift) and functions of the Touch controller, participants took a few minutes to explore virtual NYC and asked to complete two tasks.

In the first task, participants were dropped to one of several pre-selected locations in NYC, and asked to identify the neighborhood they were in.
These locations were chosen such that there was no direct sight to important city landmarks. This ensured that the user could not locate themselves easily.
Participants were then asked to navigate to one of several landmarks (e.g. Empire State Building, Chrysler Building), which they had indicated they were familiar with.

Following this, to better understand the utility of Rama mode for analysis, we simulated a real-world like analysis setting wherein participants were shown two scenarios that added new construction to NYC under 2 interpretations of zoning laws. Scenario 1 depicted 161 additional `tall' buildings, sites in NYC that could potentially accommodate a residential skyscraper of 182m or taller with a width to height ratio of 1:12. Scenario 2 depicted 171 additional `supertall' buildings, 300m or taller, with a width to height ratio of 1:20, which is considered the constructability limit of how skinny a building can be. The buildings in Scenario 2 are thinner than buildings in Scenario 1.
Participants were then tasked to navigate through the two scenarios and asked how each version of the city affected their sense of ``enclosure", and how visually dense different regions of the city appeared.
For the remainder of the paper, we use flat mode to denote the state of the system in which the Rama mode is turned off.

\subsection{Results and Discussion}
We examined participants' comments during all tasks to determine common cues and strategies they used.
Participants’ comments were transcribed, and an open coding approach was applied to responses for each task~\cite{doi:10.1002/9781405165518.wbeosg070.pub2}. Illustrative quotes have been lightly edited to remove extraneous or repeated words, without changing the semantics of the sentence. 
Based on participants' feedback during these tasks, we could classify comments into four main themes, discussed next.

\myparagraph{Localization.}
In the first task, when asked to identify the locations they were in, the visual cues mentioned by the participants can be broadly classified into the following four types:

\begin{itemize}
\item Non-target landmarks: The most common cue referenced were landmarks other than the targeted landmark (63\% of responses). These included parks (especially Central Park and Bryant Park), buildings (Bank of America Tower, Madison Square Garden, and 432 Park Avenue), and the Brooklyn Bridge (P10: \emph{Once I see the park, I'll get my bearings.}).

\item Building shape: The target landmark's shape or form was a cue mentioned by 15\% of the users. They mentioned scanning an area for the target landmark's shape, especially the shape of distinctive features, e.g. the Chrysler Building's distinctive crown (P5: \emph{I knew it had the sort of distinctive Art Deco stepping, so it wasn't too hard to find}).

\item Address or named street: Participants also referenced specific intersections or street names when evaluating their position in 15\% of responses. These references were often made in conjunction with a non-target landmark, refining the participant's position. Since no street signs or labels were used in the model, the precise numbers may have been based on existing knowledge of landmarks and NYC's street grid, estimating distance by counting blocks (P3: \emph{I'm somewhere near Times Square -- 44th street?}).

\item Cardinal directions: Participants also mentioned cardinal directions in 6\% of comments made while navigating (P5: \emph{I knew it was North}).
\end{itemize}
These cues show the importance of seeing landmarks in the surroundings for localization and navigation. Unfortunately, this is usually difficult due to occlusion when using a traditional navigation interface (flat mode).

\myparagraph{Navigation.}
There were two main strategies used to overcome occlusion for localization and navigation, namely increasing altitude and using the Rama mode.
In the flat mode the only option to overcome occlusion was to move the camera to higher altitudes in order to navigate.
While in the Rama mode, participants had the option to use either (or both) altitude and Rama mode's cylindrical projection. 
To understand the use of these strategies we recorded the participant's comments highlighting the strategies they used to navigate throughout the session. 

A common theme was that although the "bird's eye view" available at higher altitude was familiar, the Rama mode was efficient, requiring fewer interactions with the system to feel oriented (P9: \emph{So, you skip one step - you don't have to go up [...] you don't have to go up and then go down, and so it's more efficient.}; P3: \emph{certainly it's easier for you to go between~places.}). This indicates how \oursystem satisfies \textbf{R2}.

Participants also articulated that the Rama mode not only allowed them to maintain their current position, but also view and orient themselves using distant parts of the map. Being able to ``keep their place" was appreciated while navigating (P1: \emph{I have no idea where any parks are [in flat mode], but in Rama mode things that are typically horizontal become vertical, and it's kind of like a blend between the two}; P8: \emph{[In] a 3D map or a 2D map, you usually zoom out to get a wider view before zooming back in to your target. Whereas this [...] it was less necessary, because I could just navigate at whatever level I was at}; P11: \emph{That navigation was super satisfying, where it aligned me with the street so I could go straight down [the blocks].}). This highlights how \oursystem satisfies \textbf{R1}.
Three participants highlighted how simple the controls were (P9: \emph{The controls are pretty intuitive}), supporting \textbf{R3}.
To follow up on these comments, we performed a quantitative user study (Section~\ref{sec:new-user-study}) with a larger set of participants to compare height usage and task completion times when using \oursystem against existing navigation approaches.

While navigating far distances may have felt satisfying and efficient, participants trying to refine their position within a few blocks found occlusion from nearby buildings was still a problem when using the Rama mode (P8: \emph{And now the Public Library is a little hard to see because it was close, and it didn't curl up.}). One participant offered the idea of making buildings transparent while in the Rama mode, but also predicted this would come with a cost in terms of visual complexity (P1: \emph{I think it still could be useful if  you switch to this mode if the buildings become transparent, because then I could really truly see everything, but I understand that's kind of visually complex, if all that's translucent.})

\myparagraph{Real-world application.}
Overall, the users reported a very positive experience when using \oursystem and envisioned their use in their daily jobs (10 of the 11 users mentioned that UrbanRama was useful or helpful).
For example, the ability to look and move within an unlimited number of perspectives in virtual reality, rather than a few selected 2D images, was seen as a helpful way to explore options, both for expert designers and potentially for more general audiences.
P10 described how the entire VR experience differs from the typical narrative of a presentation of different building scenarios:
\emph{
You kind of have to keep all of that in your head as someone hearing the presentation, whereas this is a more immersive way of just exploring the scenarios [...] You can navigate, you can see things at scale, and in this case you're actually warping the landscape to see things, get a different kind of representation level. It's really useful. I think it's a really powerful storytelling tool also}.
P5 also reflected on how this experience differed from the usual workflow, \emph{Just the experience of being on the ground and toggling through the different options and being able to bring that qualitative aspect to it, as opposed to the quantitative which is a lot of what we do, I think it's really powerful; and it's really impressive to be able to set up different points where we can go and experience that, and compare and contrast}. 

\begin{figure*}[t]
	\centering
	\begin{minipage}[b]{0.49\linewidth}
	\centering
	\includegraphics[width=0.49\linewidth]{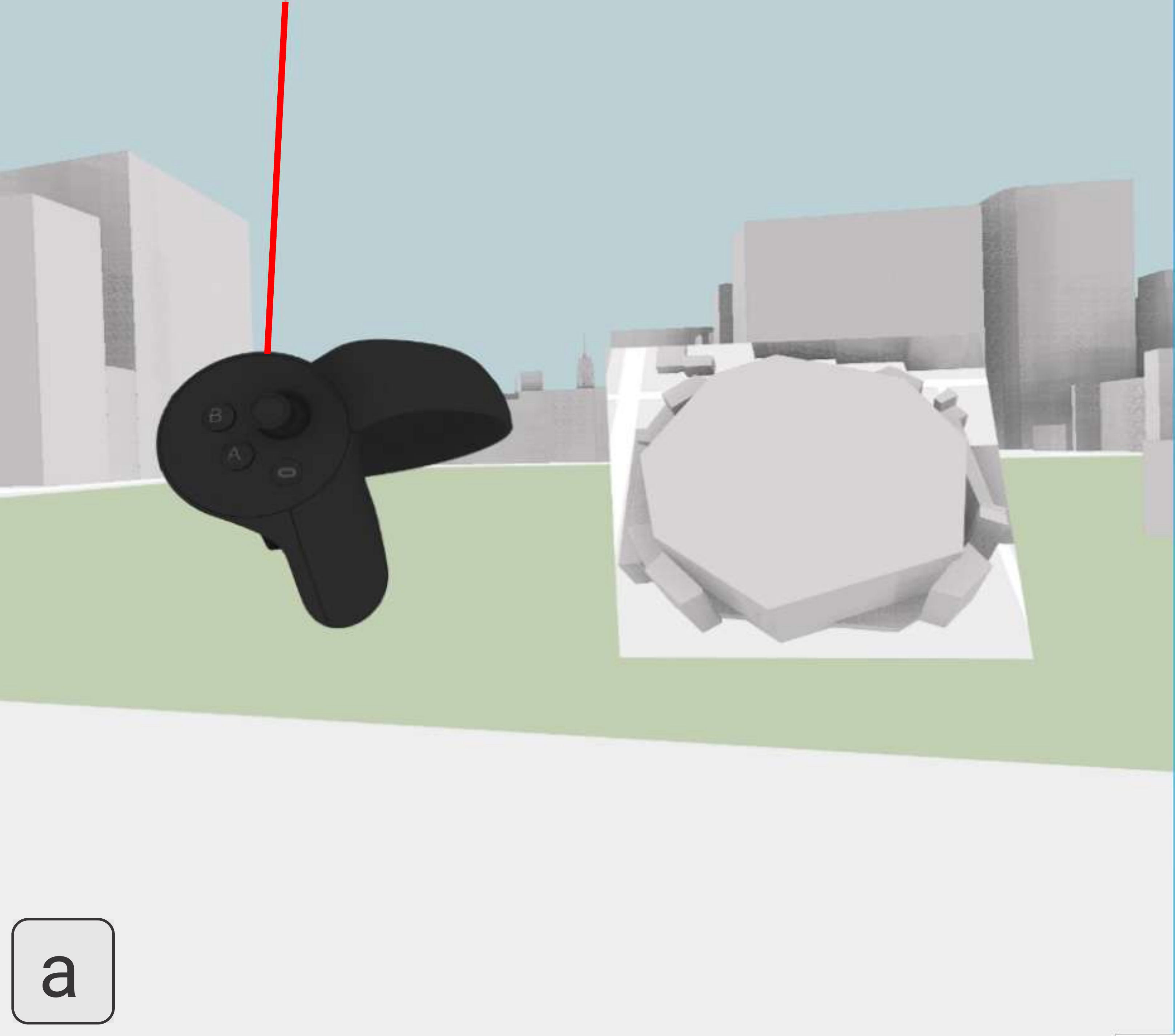}
	\includegraphics[width=0.49\linewidth]{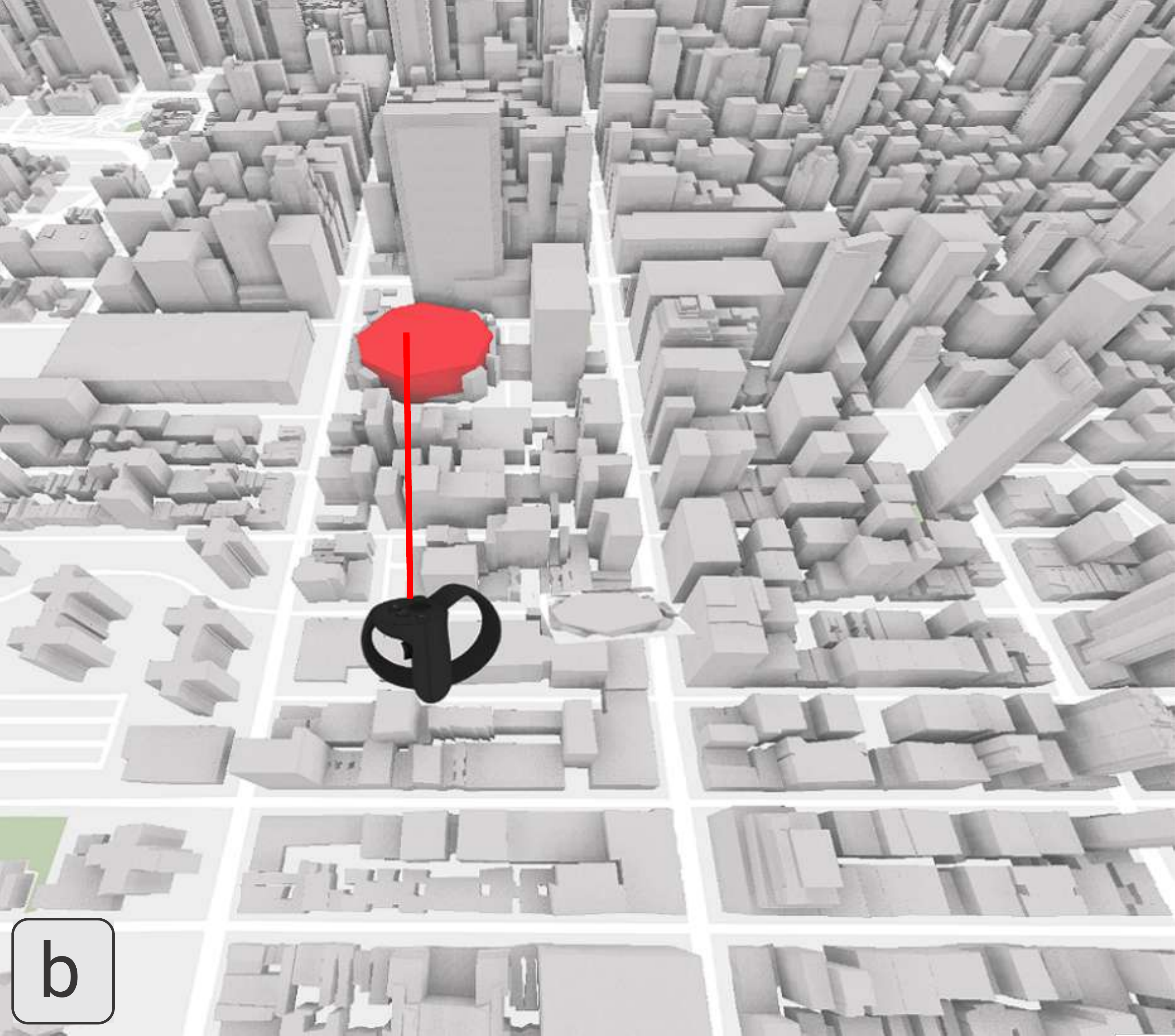}
	\caption{Task 2. (a)~User is given an image showing a landmark which is shown next to the controller. (b)~Selecting the landmark in the city.}
	\label{fig:task2}
	\end{minipage}
	\hspace{0.1cm}
	\begin{minipage}[b]{0.49\linewidth}
	\centering
	\includegraphics[width=0.49\linewidth]{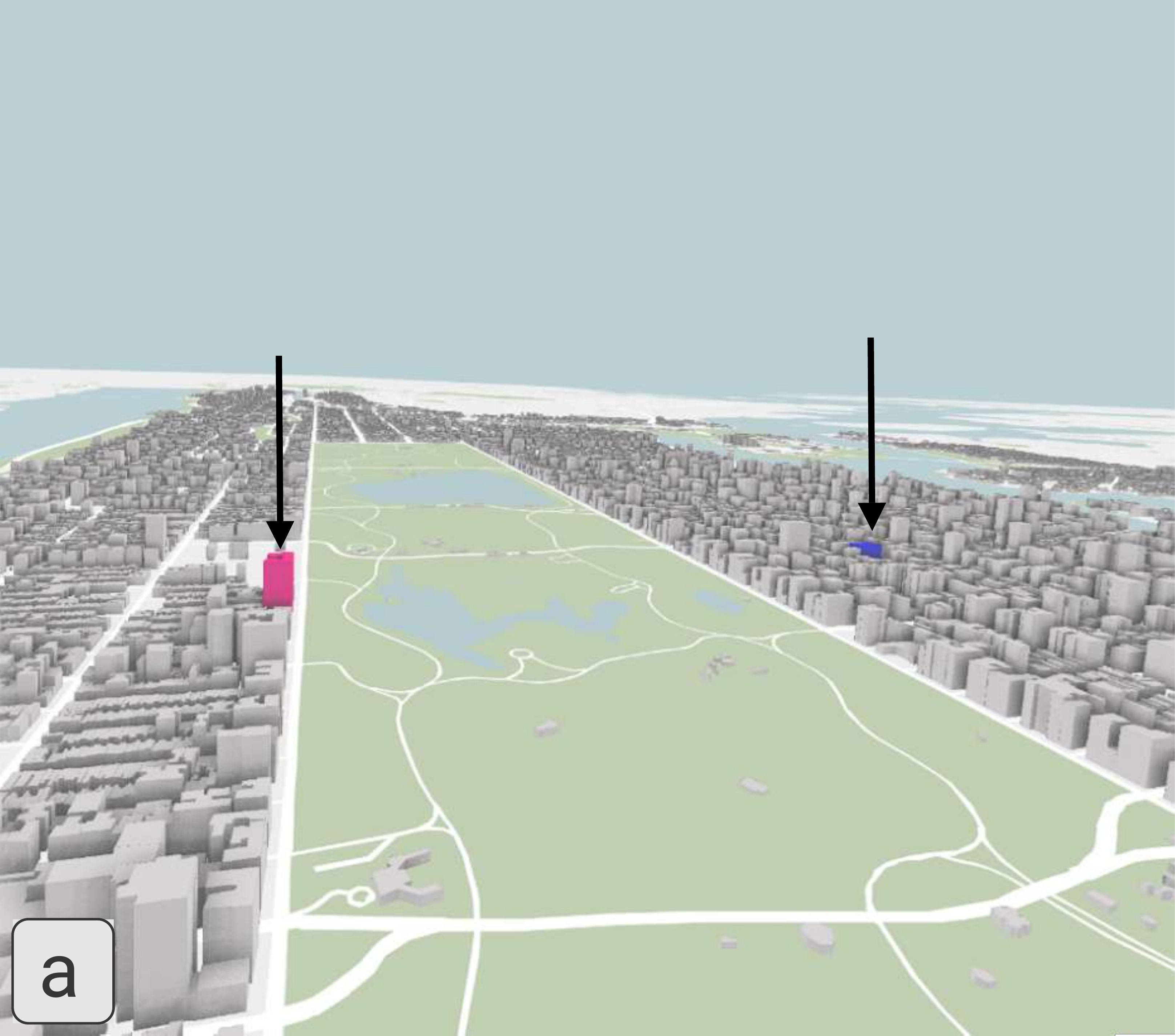}
	\includegraphics[width=0.49\linewidth]{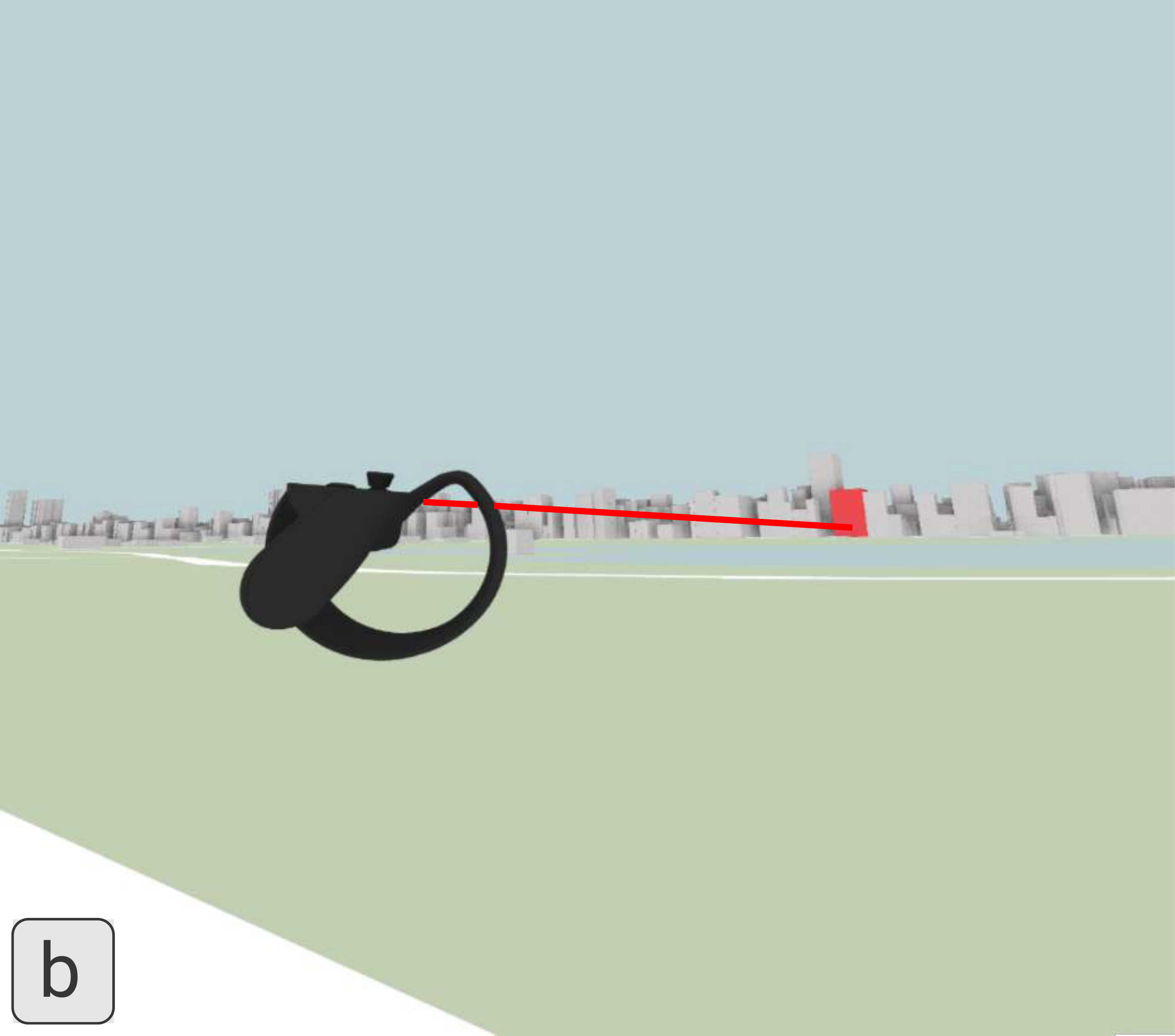}
	\caption{Task 3. (a)~The two buildings in flat mode. (b)~Pointing towards the perceived position of the blue building from near the pink building.}
	\label{fig:task3}
	\end{minipage}
\end{figure*}

\myparagraph{Global and local contexts.}
Furthermore, participants commented on UrbanRama as a unique perspective providing important context, as in P6: \emph{It puts you in a bigger context}. P9: \emph{I like that [Rama mode] because I can get the whole city}. Some participants gave more detail about specific advantages of Rama mode over flat map visualization. 
For example, P1, P5 and P7 emphasized the benefits of integrating two kinds of information into one experience. P1 mentioned that \emph{you can view both perspectives - experiential view at the ground level, and a diagrammatic view of the city beyond it, at the same time. Where usually you can only either see a street level view, or see kind of a massing model, in this case you're able to do both, which is kind of a view type that hasn't really existed before}. P5 mentioned that \emph{You're basically just doing one interface where you would get -- your attention isn't distracted from like looking down at the map and up at the 3D into toggling between the two. You've got everything all at once}. P7 said \emph{I love that you have this bird's eye view at things further away}. This again indicates \oursystem satisfies \textbf{R1}.

This study also allowed us to refine our interface and interactions. For example, several users mentioned the fact that having a constant time for the fly through (speed increased when traveling longer distances) made them uncomfortable. This resulted in making the time proportional to the distance in the final interface as described in the next section. Similarly, a threshold was used in the final interface to change the cylindrical axis to overcome the difficulty of selecting faraway buildings for navigation. 

The study also generated a number of questions/hypotheses to further explore in the quantitative study.
As presented in the Section~\ref{sec:new-user-study}, we analyze users' performances in visualizing local and global perspectives through height changes, and evaluate the effect of using Rama mode (on performance as well as perception) by looking at navigation time, angle orientation and distance judgement.

\section{The Rama Mode Final Design}
\label{sec:system}
We now discuss the final design of our system that resulted from the iterative design process.
Amongst the main differences from the previous version, we highlight the capability to fix the cylindrical axis when in Rama mode; The initial height at start of the fly through is kept during the entire movement; and the time for the fly through transition was proportional to the travelled distance instead of constant.
The supported interactions are the following:

\myparagraph{Change altitude.} 
Pushing up/down the thumbstick users can move to a higher/lower altitude.
There are 5 preset altitudes available set at heights of
5m (street level), 100m, 500m, 1km, and 2km respectively.
Moving up or down will move the user to one of these presets.

\myparagraph{Move forward.} 
The B button is used to move forward without changing the altitude. 
The direction of the headset, restricted to the horizontal plane, is used as the direction of movement.
The forward movement is also restricted to not allow the user to move through buildings.

\myparagraph{Enter / Exit Rama mode.} 
The A button is used to enter and exit Rama mode. 
In order to preserve context during this switch,
we animate this transition between the two modes.
When user goes from regular mode to Rama mode, the initial diameter of the cylinder 
used for the deformation is set to 10000km (which creates almost no deformation).
We then gradually reduce this diameter in each frame
until it becomes 5km.
As seen in the accompanying video, this will
result in an animation where the city gradually bends towards the user.
The time period of the animation is fixed at 3 seconds,
and the intermediate diameter for the transition is
computed using logarithmic interpolation. 
By using the logarithmic value, we can make user feel that the bending goes linearly since the change of diameter looks more subtle as the diameter becomes larger.

The reverse transition is applied when exiting Rama mode, that is,
the diameter of the cylinder is increased from 5km
to 10000km, before rendering the scene in the regular mode.
Note that all interactions on the controller are temporarily disabled until the transition is complete.

\myparagraph{Fly to a selected building or point.} 
In addition to the button controls, the right controller also acts as a laser pointer. 
We use this to allow users to point to locations of interest.
For example, pointing to a particular building selects and highlights it. 
Pointing to a location or building and squeezing the trigger will move the user to the desired target. 
In case this is a building,
then the user is moved to a point near that building.
There are two ways in which to accomplish this movement of the user, namely, "teleport"  and "fly-through".
We evaluated both methods and found that the expert users assessed the teleportation strategy disorienting and caused them to lose the spatial context where they were located.
So, we finally chose to fly the user to the destination through a smooth transition.
Based on users' feedback (see Sections~\ref{sec:iterative-design} and \ref{sec:user-study}) 
we moved the camera to the destination location but retaining its initial altitude.
To make the transition look natural and smooth, we use a sine easing function $t' = (1 - \cos(\pi \cdot t))$ to interpolate the user camera from the start point to the destination.
The travel time is proportional to square root of the travel distance,
and all interactions on the controller are temporarily disabled during the travel.
These transition settings are also applied when user changes the altitude.
The collision avoidance will be temporarily disabled during the fly through, {\it i.e.} 
if there are buildings between the start and final points, the user will fly through them. 
Also, if the Rama mode is active at the start of this movement, then this mode is turned off when the user reaches the destination. 
In addition, during this fly through, both the center and the orientation of the cylinder are fixed.
These were again design choices that were chosen based on feedback from our collaborators after trying out several options.

\myparagraph{Being in Rama mode.}
When in Rama mode, the orientation of the cylinder used for the
deformation follows the headset direction changing
whenever the user turns around or looks around. 
Thus, the city is always ``bent" towards the user.
Note that the headset has high sensitivity capturing even minute 
head movements. 
In our initial evaluations, some of the users 
found this distracting  when using the system. 
They also found this was affecting 
them when trying to select a building / point of interest to teleport to.
To avoid this issue, we set a small threshold for the head movement so that
a change in cylinder orientation is triggered only when the headset movement
(computed as the angular velocity and angular displacement between eye direction
and current cylinder orientation) crosses this threshold.

Optionally, we also allow the user to pause the cylindrical axis for this
deformation by squeezing the grip of the controller.
This allows the user to look at different regions in the city
after fixing the deformation. The continuous deformation mode
can then be resumed by again squeezing the grip.
In case the resumption changes the 
axis of the cylinder, then a smooth transition is 
used to move to the new deformation.
The $x$ and $y$ coordinates of the cylinder center also follow the position of user when user presses the B button to move forward.

\subsection{Implementation}
\label{sec:implementation}

We implemented the prototype as a browser-based application using JavaScript, WebGL~2.0 and WebVR.
The Rama mode deformation is implemented by using the vertex shader.
We used data from Open Street Maps~\cite{OSM}~(OSM) to create the virtual city.
This included the polygonal boundaries of the city including the water bodies, parks, and road network,
as well as the 3D mesh corresponding to the buildings in the city.
We use Screen Space Ambient Occlusion~\cite{SSAO}~(SSAO) for shading the buildings. 
When the user is in Rama mode, the positions in the deformed space are used to calculate the ambient occlusion instead of the positions in world space.
\section{Quantitative User Study}
\label{sec:new-user-study}

\begin{figure}[t]
	\centering
	\includegraphics[width=0.45\linewidth]{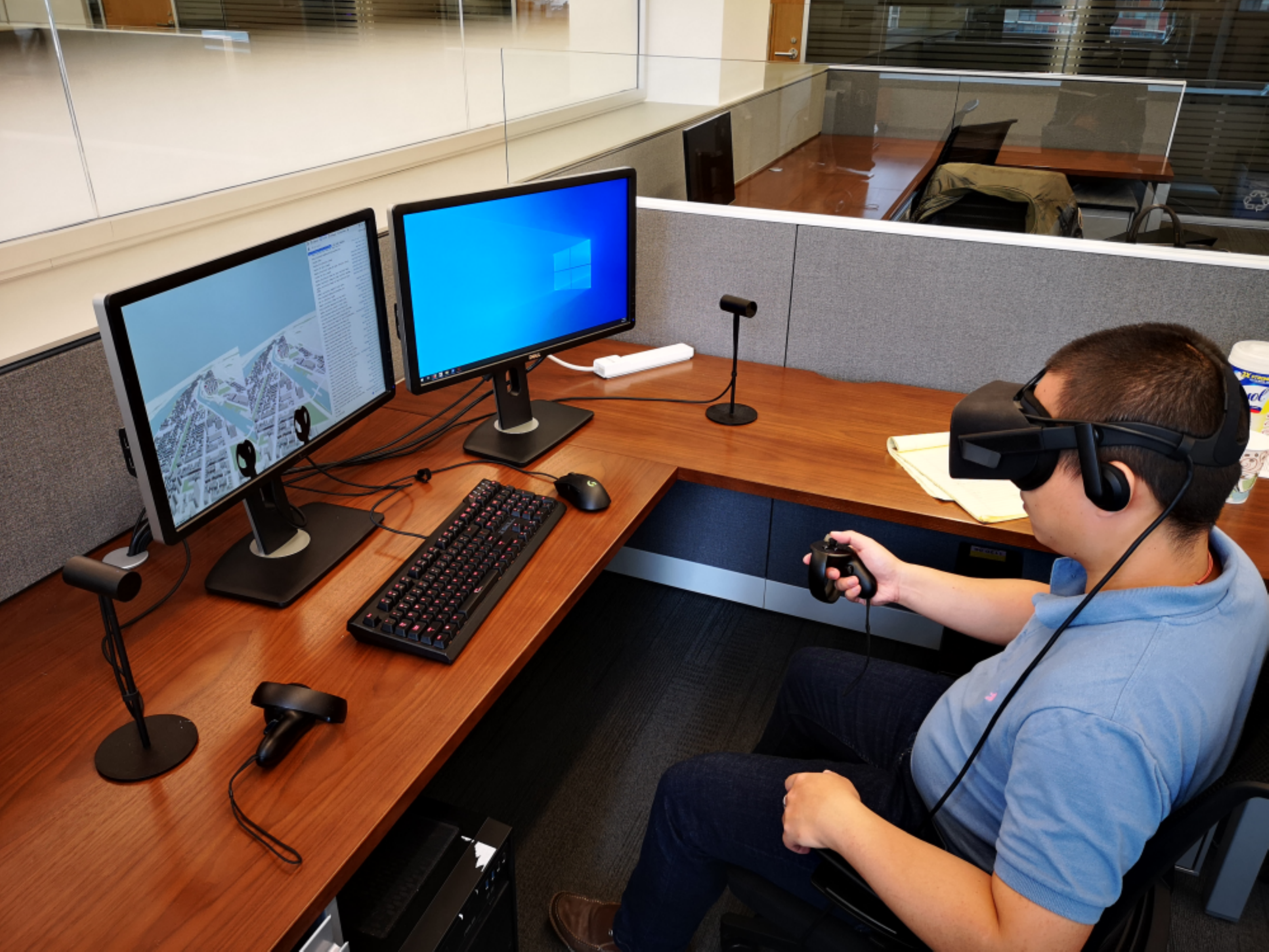}
	\includegraphics[width=0.45\linewidth]{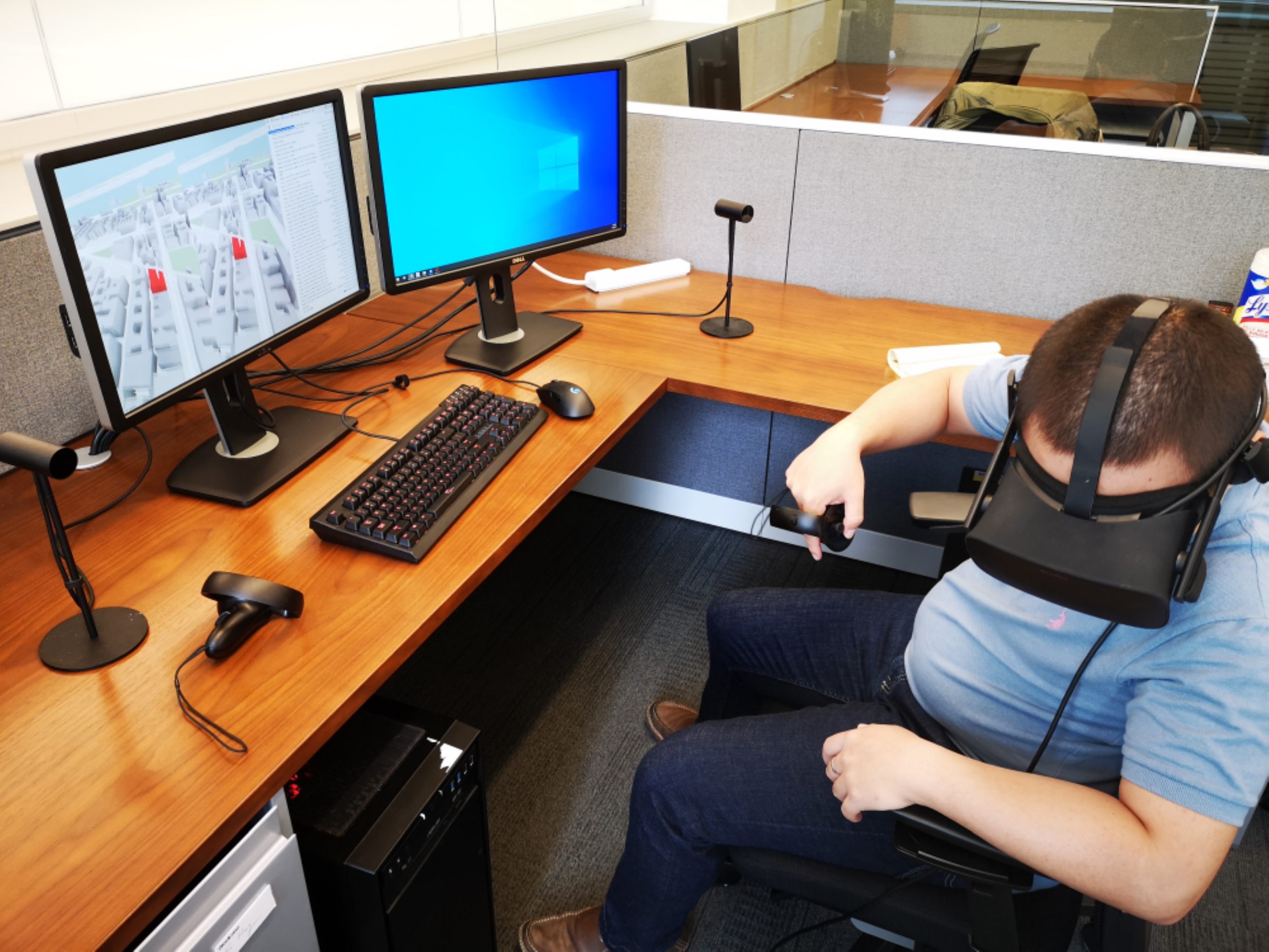}
	\caption{The setting of the quantitative user study. Both images were captured from our user study.}
	\label{fig:user_study}
\end{figure}

\begin{figure*}[t]
	\centering
	\includegraphics[width=0.285\linewidth]{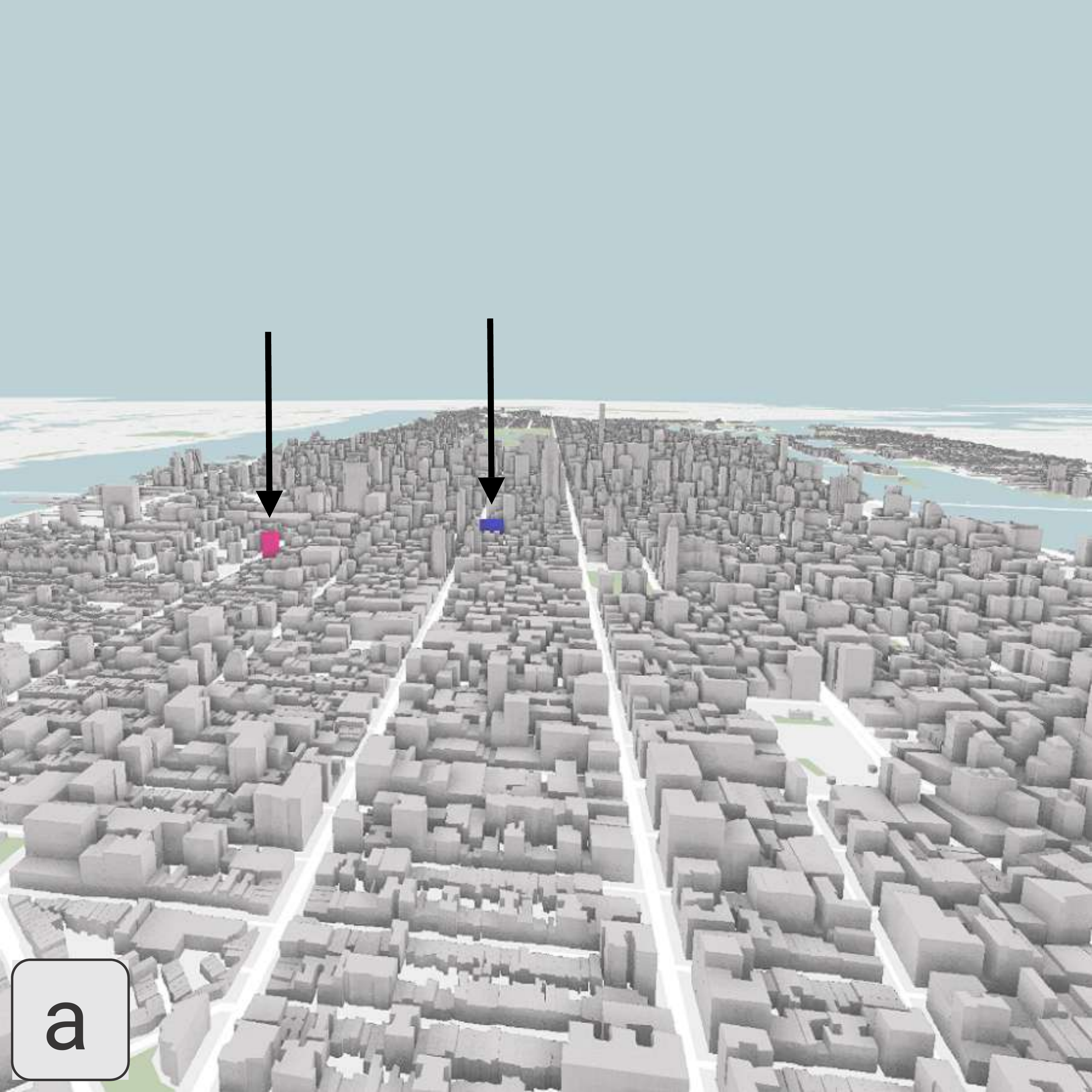}\hspace{0.03\linewidth}
	\includegraphics[width=0.285\linewidth]{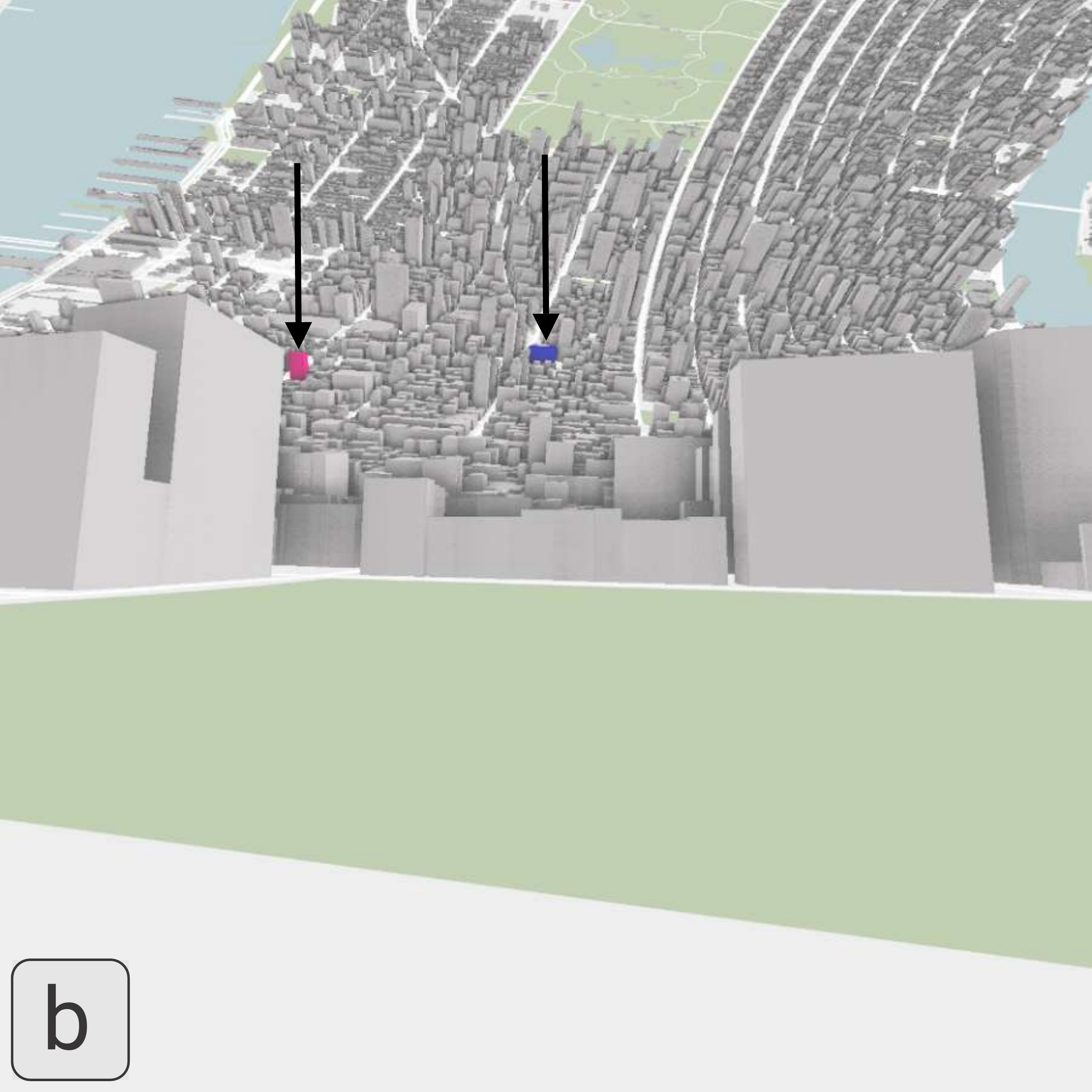}\hspace{0.03\linewidth}
	\includegraphics[width=0.285\linewidth]{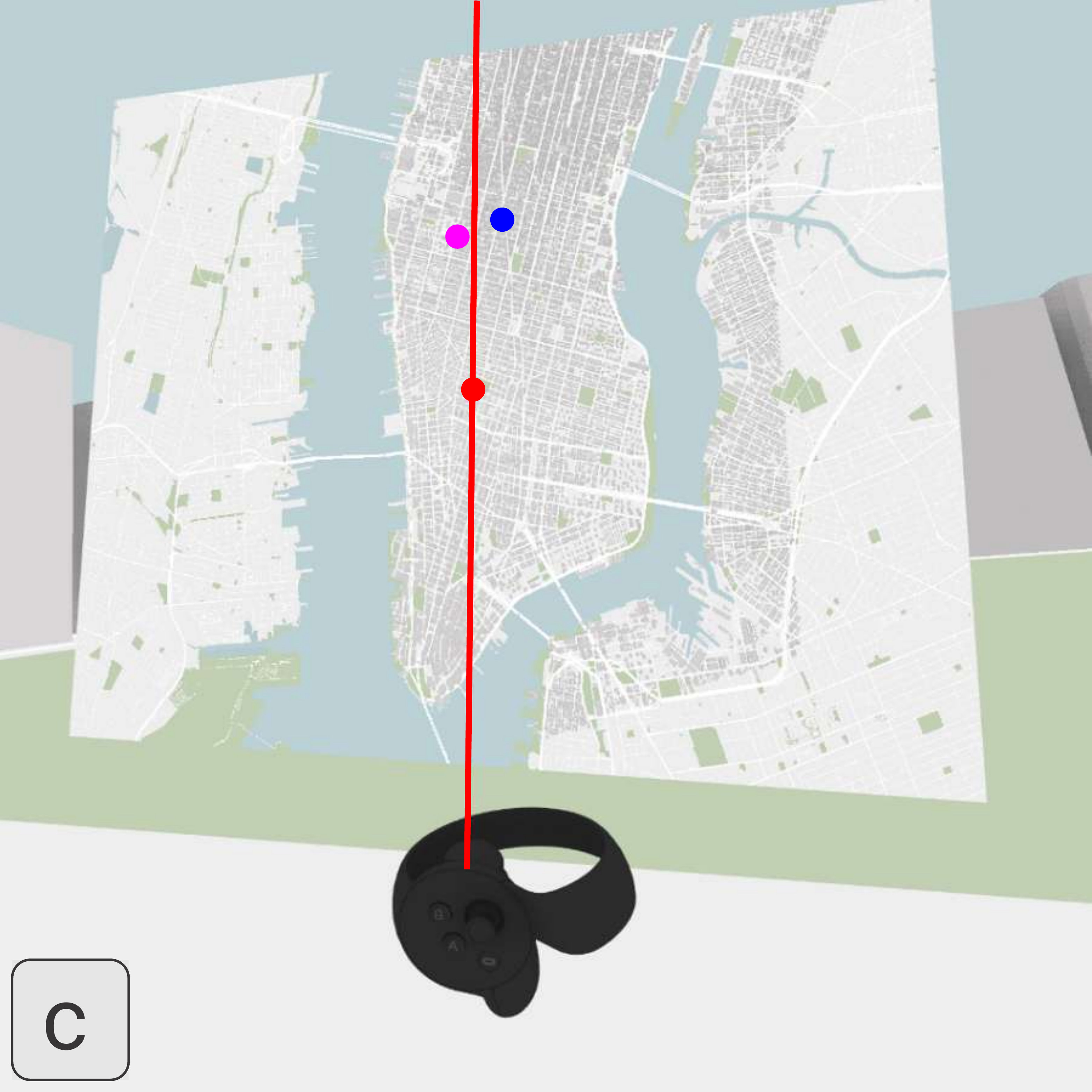}
	\caption{Different modes tested in our quantitative user study: flat mode (a), Rama mode (b) and minimap mode (c). The three figures highlight the distance comparison in Task 4.}
	\label{fig:task4}
\end{figure*}

Using the experience and feedback from the qualitative evaluation, we performed a quantitative one to assess the effectiveness of our proposal. 
Our goal was to measure the impact of \oursystem on common urban environment exploration tasks, as well as to measure the simulator sickness.
To do so, we compared \oursystem against two alternatives that are common in current systems: flat mode and a minimap mode~\cite{minimap}. Fig.~\ref{fig:task4} shows the three modes.
The minimap mode has the same interactions as the flat mode, in addition to a minimap showing a bird's eye view of current location on top of the controller model, as shown in Fig.~\ref{fig:task4}(c).
The minimap is a ``forward-up'' minimap such that the north of the map always follows the orientation of the headset.
The study was performed over the course of a week, with each user session taking from 15 to 30 minutes to complete. 
The setting of the study is shown in Fig.~\ref{fig:user_study}.

\myparagraph{Participants}
There were 30 participants, 27 male and 3 female, between the ages of 22 and 41, and they were recruited from the computer science, electronic engineering and urban science departments at New York University; 20 of them were graduate students, 5 MSc graduates, and 5 PhD graduates.
None of the participants had previously used the system before the study.
The participants were randomly assigned to one of the following 3 groups: flat (which was the control group), minimap and Rama mode, so that each group had 10 participants. 
25 participants had prior experience with video games and 14 participants had prior experience with VR (5 in the flat mode group, 4 in the Rama mode group and 5 in the minimap group). 

\subsection{Methods and Data Collection}

Before the start of the session, each participant was asked to complete a pre-task Simulator Sickness Questionnaire~(SSQ)~\cite{SSQ}. We also asked the participants to give a score between 0 (no pain) and 3 (severe pain), if there was any symptom of neck pain.

The actual study session consisted of the four tasks,
all of them were tested in different cities to control for city layout and also for possible biases due to prior knowledge of the city. 
New York was selected for cities with rectangular grid layout, and London and Tokyo were selected for cities with irregular layout.
The first two tasks focused on measuring the effectiveness of navigation (\textbf{R1,R2}), while the last two tasks were used to measure if the warping performed by \oursystem had any adverse effects with respect to perception and usability.

\myparagraph{Task 1 - Navigation.}
The participants were asked to move to 5 highlighted buildings in the given order---the first building was initially highlighted, and the next one highlighted when the user completes the move to the previously highlighted building. 
The move to a building was deemed to be complete when the participant is positioned on a street adjacent to that building and at street level. 
This task was repeated twice, once in NYC and once in Tokyo. 
Completion time and percentage of time spent in different altitude levels were recorded as a way to assess how the different modes supported the design goals R1 and R2.

\myparagraph{Task 2 - Search.}
The participant was shown an image of a building (next to the rendering of the controller), and was tasked with finding the building in the virtual world (Fig.~\ref{fig:task2}(a)). 
Once found, the participant was asked to point towards the building and hold the button A for 3 seconds to indicate that the building was found (Fig.~\ref{fig:task2}(b)). 
This task was repeated twice, once in NYC and once in London. 
Completion time and percentage time spent in different altitude levels were recorded as a way to assess how the different modes supported the design goals R1 and R2.

\myparagraph{Task 3 - Orientation Estimation.}
In this task, two buildings were highlighted--one in blue and another in pink, and the participant was placed at a starting position at street level such that the blue building is occluded. 
Users were restricted to limited interactions: only the altitude could be changed by users in the flat mode group; only Rama mode could be toggled in the Rama mode group; and only the minimap could be used (with the buildings highlighted in the minimap as well) in the minimap group. 
Once the scene was observed using the provided options (Fig.~\ref{fig:task3}(a)), the participant was moved to a position near the pink building and 
tasked with pointing towards the blue building (Fig.~\ref{fig:task3}(b)).
The pointing angle error was then recorded.
This task was also performed twice, once in NYC and once in~London. 

\myparagraph{Task 4 - Distance Comparison.}
Again, two buildings were first highlighted--one in blue and another in pink %
The task was then to identify the building closest to the participant's location.
The starting position was at the street level where the 2 buildings are occluded and limited interactions (same as in the previous task) were made available. 
This task is repeated 6 times, thrice in NYC and thrice in London, and number of error(s) was recorded. 
Distances to buildings were set to between 1.3 and 2.1km. In each trial the distances from the highlighted buildings to the user differ by 5\%~to~15\% of the closest distance. Fig.~\ref{fig:task4} shows the highlighted buildings in the different~modes.

Once all the tasks were completed, the participants were asked to complete a post-task SSQ, as well as score their neck pain level.

\subsection{Results}

\begin{table*}[t]
\begin{minipage}[b]{0.49\linewidth}
\centering
\caption{Percentage time spent in given altitude during Task 1 in NYC.}
\begin{tabular}{c c c c c c}
\hline
Group & 5m & 100m & 500m & 1km & 2km\\ \hline
\hline
Flat & 19\% & 18\% & 41\% & 14\% & 7\%  \\ \hline
Rama & 62\% & 19\% & 17\% & 3\% & 0\%  \\ \hline
Minimap & 27\% & 18\% & 30\% & 16\% & 8\% \\ \hline
\end{tabular}
\label{table:T1-NYC-h}
\end{minipage}
\begin{minipage}[b]{0.49\linewidth}
\centering
\caption{Percentage time spent in given altitude during Task 1 in Tokyo.}
\begin{tabular}{c c c c c c}
\hline
Group & 5m & 100m & 500m & 1km & 2km\\ \hline
\hline
Flat & 11\% & 13\% & 29\% & 33\% & 14\%  \\ \hline
Rama & 73\% & 13\% & 10\% & 4\% & 0\%  \\ \hline
Minimap & 17\% & 17\% & 30\% & 26\% & 9\% \\ \hline
\end{tabular}
\label{table:T1-TYO-h}
\end{minipage}
\end{table*}

\begin{table*}[t]
\begin{minipage}[b]{0.49\linewidth}
\centering
\caption{Percentage time spent in given altitude during Task 2 in NYC.}
\begin{tabular}{c c c c c c}
\hline
Group & 5m & 100m & 500m & 1km & 2km\\ \hline
\hline
Flat & 20\% & 3\% & 25\% & 37\% & 16\%  \\ \hline
Rama & 45\% & 12\% & 29\% & 14\% & 0\%  \\ \hline
Minimap & 13\% & 8\% & 36\% & 26\% & 17\% \\ \hline
\end{tabular}
\label{table:T2-NYC-h}
\end{minipage}
\begin{minipage}[b]{0.49\linewidth}
\centering
\caption{Percentage time spent in given altitude during Task 2 in London.}
\begin{tabular}{c c c c c c}
\hline
Group & 5m & 100m & 500m & 1km & 2km\\ \hline
\hline
Flat & 10\% & 5\% & 59\% & 25\% & 1\%  \\ \hline
Rama & 30\% & 24\% & 39\% & 6\% & 1\%  \\ \hline
Minimap & 11\% & 9\% & 54\% & 23\% & 2\% \\ \hline
\end{tabular}
\label{table:T2-LDN-h}
\end{minipage}
\end{table*}

\begin{table*}[t]
\begin{minipage}[b]{0.49\linewidth}
\centering
\caption{Total severity score.}
\begin{tabular}{c c c c}
\hline
Group & Pre $\pm$ Std & Post $\pm$ Std & Diff\\ \hline
\hline
Flat & 8.98$\pm$15.08 &  17.58$\pm$14.65 & 8.60$\pm$16.64 \\ \hline
Rama & 11.22$\pm$18.07 & 28.05$\pm$24.64 & 16.83$\pm$26.34 \\ \hline
Minimap & 9.72$\pm$9.36 & 13.09$\pm$12.12 & 3.66$\pm$12.40  \\ \hline
\end{tabular}
\label{table:TS-Score}
\end{minipage}
\begin{minipage}[b]{0.49\linewidth}
	\centering
	\caption{Total severity score of experienced (E) and inexperienced (I) VR users.}
    \begin{tabular}{c c c c}
		\hline
		Group & Pre $\pm$ Std & Post $\pm$ Std & Diff\\ \hline
		\hline
		Flat~(E) & 6.73$\pm$3.97 &  8.98$\pm$6.20 & 2.24$\pm$6.38 \\ \hline
		Flat~(I) & 11.22$\pm$4.58 &  26.18$\pm$14.87 & 14.96$\pm$14.87 \\ \hline
		Minimap~(E) & 13.46$\pm$12.00 &  15.71$\pm$10.38 & 2.24$\pm$14.87 \\ \hline
		Minimap~(I) & 5.98$\pm$4.18 & 10.47$\pm$13.86 & 4.49$\pm$10.07 \\ \hline
		Rama~(E) & 12.16$\pm$19.64 &  23.38$\pm$23.13 & 11.22$\pm$20.26 \\ \hline
		Rama~(I) & 10.60$\pm$18.84 &  31.17$\pm$27.24 & 20.57$\pm$31.00 \\ \hline
	\end{tabular}
	\label{table:TS-Score-2}
\end{minipage}
\end{table*}

\myparagraph{Task 1.}
Table~\ref{table:T1-NYC-h} and Table~\ref{table:T1-TYO-h} present the average percentage of time spent at a given altitude during Task 1 for NYC and Tokyo, respectively. 
The participants using the Rama mode spent on average 81\% of the task time in lower altitudes (5m and 100m) while completing the task in NYC, and 86\% in Tokyo. On the other hand, flat mode users spent 37\% and 24\% respectively in lower altitudes in NYC and Tokyo, and users on minimap mode spent 45\% and 34\% respectively. 

\begin{figure*}[t]
	\centering
	\includegraphics[width=1.0\linewidth]{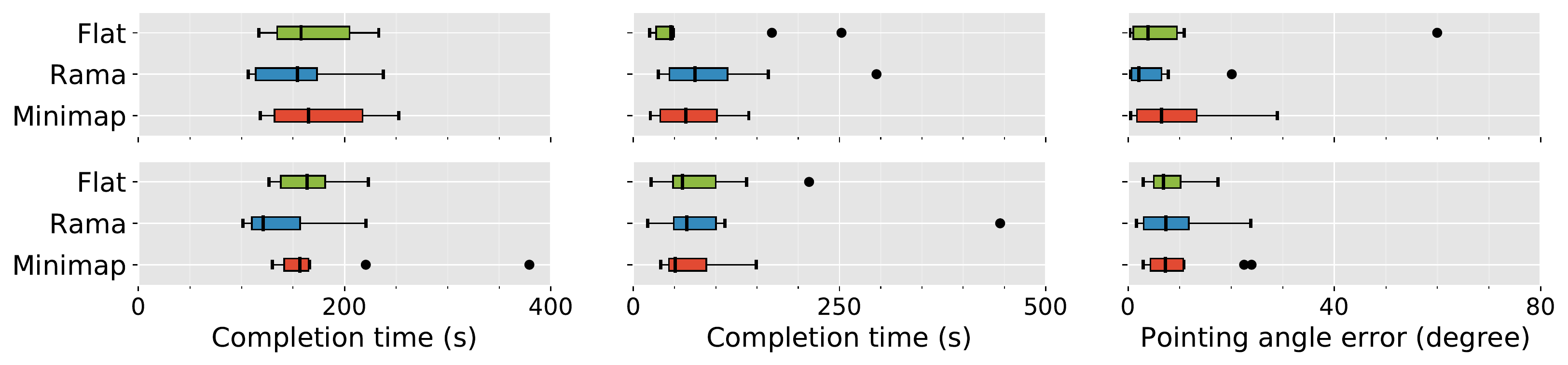}
	\caption{Left: completion time for Task 1 in NYC (top) and Tokyo (bottom), Center: completion time for Task 2 in NYC (top) and London (bottom), Right: pointing angle error for Task 3 in NYC (top) and London (bottom). Black dots represent outliers, i.e., values larger than the upper quartile by at least 1.5$\times$ the interquartile range.}
	\label{fig:tasks}
\end{figure*}

\begin{figure*}[h]
	\centering
	\includegraphics[width=0.9\linewidth]{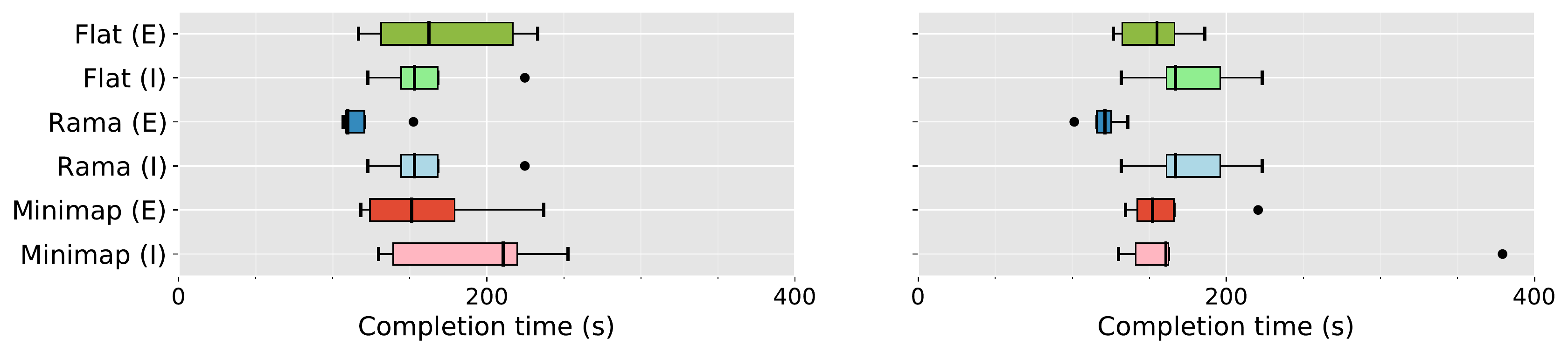}
	\caption{Task 1 completion time of experienced VR users (E) in the three groups, and inexperienced VR users (I), for NYC (left) and Tokyo (right).}
	\label{fig:plot2}
\end{figure*}

The participants using the Rama mode also switched perspectives less frequently between lower altitudes (5m and 100m) and higher altitudes (500m, 1km and 2km); on average they switched 2.1 times in NYC and 1.2 times in Tokyo. Users of the flat and minimap mode, on the other hand, went from lower altitudes to higher altitudes much more frequently, on average 6.2 and 6.0 times in NYC and 5.4 and 5.7 times in Tokyo.
Fig.~\ref{fig:tasks}(left) shows the total time taken for completing Task 1. 
While we did not find an overall statistically significant improvement, we point out that the results do indicate an improvement when using Rama mode, both in NYC (flat:$167.37 \pm 43.03$, Rama:$153.86 \pm 43.69$, minimap:$176.13 \pm 50.38$), but especially Tokyo (flat:$164.55 \pm 30.82$, Rama:$138.05 \pm 39.62$, minimap:$178.85 \pm 74.91$), where buildings are not as tall and dense as in NYC.
Furthermore, an interesting observation could be made if we segment the participants based on their declared VR experience.
As can be seen in Fig.~\ref{fig:plot2}, experienced VR users are considerably faster when using Rama mode to perform this task.

\myparagraph{Task 2.}
The results for Task 2 (Fig.~\ref{fig:tasks}(center)) present a similar picture when considering average percentage of time spent at lower altitudes: Rama mode has an advantage in maintaining users at lower altitudes (local perspective), as shown in Tables~\ref{table:T2-NYC-h} and~\ref{table:T2-LDN-h}. 

Task 2 was the most challenging task in the entire study. Since buildings were just shaded using SSAO, it was difficult to distinguish shapes of far away buildings.
In NYC, users in the flat mode group completed this task in less time ($70.99 \pm 76.72$) than users in the Rama group ($97.70\pm81.28$), but similar to the minimap group ($69.36 \pm 42.36$).
When considering London, the Rama group also took more time ($103.86 \pm 123.11$), when compared to the other two groups (flat: $79.83 \pm 58.85$, minimap: $68.10 \pm 38.79$).

We note that, similar to Task~1, VR experience played an important role in the results.
Inexperienced users in the Rama group took more time to complete the task in NYC ($130.32 \pm 90.75$) and London ($144.21 \pm 135.84$) when compared to the flat mode (NYC: $111.99 \pm 89.41$, London: $89.16\pm48.83$) and the minimap group (NYC: $87.45\pm44.91$, London: $75.39\pm42.08$).
Experienced users in the Rama group took more time in NYC ($48.76\pm27.78$) when compared to the flat group ($30.00\pm8.61$), but similar to the minimap group ($51.26\pm31.00$).
On the other hand, in London, experienced users in the Rama group completed the task in less time ($43.32\pm20.32$) when compared to the other two groups (flat: $70.49\pm12.60$, minimap: $60.80\pm30.74$).

\myparagraph{Task 3.}
Fig.~\ref{fig:tasks}(right) shows the pointing angle error for the three groups. In NYC, the error for the group in Rama mode ($4.48 \pm 6.16$) was smaller than the other two groups (flat: $9.90 \pm 18.03$, minimap: $9.25 \pm 9.59$), while the errors obtained from the London groups were much closer across the three modes (flat: $7.88 \pm 18.03$, Rama: $8.91 \pm 7.56$, minimap: $9.74 \pm 7.65$).
We believe this can be attributed to the grid layout of the streets in NYC, that could serve as a visual guide while completing the task. 
The result indicates that enabling Rama mode does not affect the perception of orientation.

\myparagraph{Task 4.}
In Task 4, users in the minimap group, as expected, correctly selected the nearest building in all scenarios. This was expected, since the minimap clearly indicates the position of the buildings in a  map. In the flat mode, out of 60 tests there were two mistakes by two different users (3.3\% overall error). In the Rama mode, there were three mistakes by three different users (5\%). This small difference between the two groups suggests that the Rama mode does not impact judgment of distance.

\begin{figure}[h]
	\centering
	\includegraphics[width=0.9\linewidth]{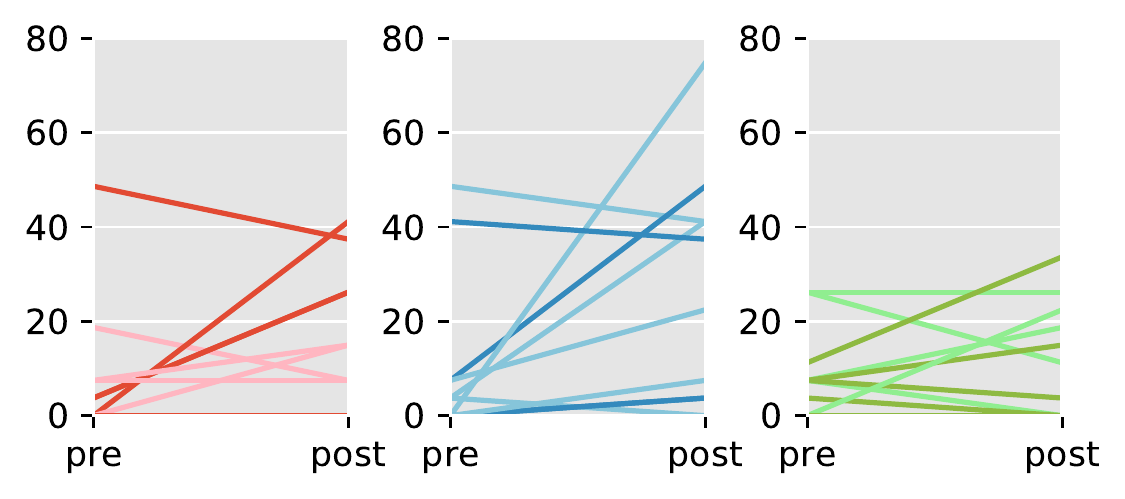}
	\caption{Pre and post SSQ scores for minimap (left), Rama mode (middle), flat (right). Here we use the same colors as Fig.~\ref{fig:plot2}, i.e., lighter shades of red, blue, and green indicate inexperienced users, and darker shades indicate experienced users.}
	\label{fig:plot3}
\end{figure}

\myparagraph{Simulator sickness.}
Even though the SSQ score of UrbanRama is greater than the other two modes (Table~\ref{table:TS-Score}), it is still below the thresholds used to consider the start of severe simulator sickness (60 by~\cite{vr2004,10.3389/fpsyg.2019.00158} and 70 by~\cite{IEEEVR-automatic,IEEEVR-occlusion-study,doi:10.1177/154193129804202101}). Similar systems also reported similar SSQ scores~\cite{IEEEVR-automatic,Magic-Carpet}. The increase between pre and post SSQ, however, indicates that slight simulator sickness is still a factor to be considered in both flat and Rama mode (Fig.~\ref{fig:plot3}). Furthermore, by segmenting the participants based on their declared VR experience, we can find that experienced users had a smaller increase of simulator sickness than inexperienced users, especially for flat mode and Rama mode (Table~\ref{table:TS-Score-2}). 
In the Rama mode group, one inexperienced user in particular reported a pre-SSQ score of 0, followed by a post-SSQ score of 74.8 (outlier in Fig.~\ref{fig:plot3}(middle)).
We hypothesize that users’ inexperience may also be a cause of the increased SSQ.

\subsection{Summary Discussion}

Overall, the results presented in the previous sections suggest that the final design of Urban Rama achieved its original design goals.
More clearly, we could observe a  significant time spent in lower altitudes and a small number of perspective changes between altitudes
compared to the other modes in Tasks 1 and 2.
This indicates that our proposal provides the global perspective needed to complete the task, while maintaining the local context, satisfying \textbf{R1}.

Concerning R2, the completion times for Tasks 1 and 2 suggest that participants were able to find and navigate to points of interest efficiently using \oursystem. 
This can be seen clearly for Task 1, but was not strongly supported for Task 2. 
However, the timings for Task 2 are comparable with the other modes, suggesting that the participants were able to finish this task while in a similar time, but preserving the overall context and performing less altitude changes. 
Another important thing to notice is that our results suggest prior 
VR experience had a big impact on how the participants approached the use of \oursystem.
Again, this pattern is clearer in Task 1, but was also present in Task 2.
While the samples are too small to perform actual statistical analysis, we hypothesize that due to not being familiar with the VR environment, inexperienced participants could not take full advantage of Rama mode.
In addition, participants with prior VR exposure seem to more easily use the Rama mode capabilities to finish the search and navigation tasks faster than using the other modes.
One hypothesis that came out of these observations is that once these users get accustomed to the VR setup, the results would show an improvement in navigation times using our approach.
Furthermore, this improvement would show faster completion times compared to the other modes.
We plan to perform a larger user study to investigate this hypothesis.

Usability was also a big concern when designing \oursystem. The overall feedback is that participants were able to quickly learn to use the commands present in Rama mode. This shows that our final design achieved R3. 
In addition, we were aware that using a spatial deformation strategy 
could result perceptual or motion sickness problems during the use of our approach.
However, the results for Tasks 3 and 4 suggest that the deformation strategy used in the Rama mode did not impose an impact on the perception of orientation and distance estimation. Also, simulator sickness was still in acceptable levels.
Finally, due to the deformation that is performed by the Rama mode, one would expect additional strain on the neck if users tend to look up frequently. However, we noticed that only two participants reported an increase in the neck pain score in the questionnaire: one in the minimap group and the other in the Rama group. Their reported score increased from 0 (no pain) to 1 (slight pain). All other participants, from all groups, either left the neck pain score unchanged or decreased it. This indicates that neck pain is not a major concern when using Rama mode.

\section{Conclusions and Future Work}

In this paper, we presented \oursystem, a new way to visually explore urban landscapes in VR that provides users with both a local as well as a global perspective of the virtual city (\textbf{R1}). 
User studies (Sections~\ref{sec:user-study} and \ref{sec:new-user-study}) demonstrated that it succeeded in this endeavor using a small number of controls (\textbf{R3}) and without hampering the user's perception of distance, orientation, and without degrading the navigation experience (\textbf{R2}).
Although these experimental studies were exploratory and we were only able to observe trends, we believe that the emerging themes offer a useful foundation for further work in this area. In future work, we plan to further explore the impact of deformations in general in urban VR navigation. This would demand a more comprehensive user study, with more participants to extract significant results, as opposed to trends.
A factor we would like to consider in a future user study is how well the user knows the city, given that this might have played a role in our search task. Moreover, we would also like to study the impact of city size on navigation and search tasks.

The results of both the user studies also pointed out some limitations and directions for future development.
While users usually are able to see far away locations using \oursystem, its utility is still restricted when located in a dense neighborhood, especially at street level.
While changing the altitude of the user mitigates such problems, other solutions such as adding an ``X-ray" or see through mode \cite{Avery2009Xray,Dey2011xray} might also be helpful. 
For example, selectively making a subset of the buildings transparent so that user can see the occluded landmarks or even overlaying outlines of major landmarks over closer buildings could help preserve the context for the user.

In \oursystem, far-away buildings are slightly deformed, since we want to preserve the distances between buildings. In our experience this deformation is visually negligible. However, for application scenarios in which this deformation is not acceptable we plan to investigate approaches that preserve shape in the eventual case that shape preservation is more important than preserving distances. One such approach is to unitize the void space among buildings~\cite{tong2017glyph}.
In a future iteration of the project, we intend to investigate the use of \oursystem for visual data analytics commonly used by architects and urban designers. 
For example, several users commented on their desire to study how shadows would change over the course of a day or between seasons. 
Representations of shadow accumulation metrics are an important source of information for experts like these users~\cite{8283638}, but being present at street level in a virtual environment offers more insight for experts and laypeople alike.
Similar interest and concerns were expressed about assessing visibility and how it changes with new urban developments. 
Similarly, the integration of other data sources such as real-time data streams related to noise, traffic, and energy use is of great interest for urban planning professionals.
This raises the need for novel navigation capabilities such as navigating to areas based on specific metric constraints. 
We believe that such analysis scenarios can greatly benefit from \oursystem and are interesting avenues for future research. 
Another direction of work that we intended to explore in the future is to support collaborative exploration of the urban environment by multiple users in the same virtual space simultaneously.
We believe the experience of using \oursystem would be enhanced by having multiple points of view.
Developing interaction and visual metaphors that allow users to communicate and coordinate in these environments are the main research tasks in this case.


%




\section*{Acknowledgment}
We would like to thank our colleagues from Kohn Pedersen Fox and New York University for their help in this research.
This work was supported in part by: NSF awards CNS-1229185, CCF-1533564, CNS-1544753, CNS-1730396, CNS-1828576, CNS-1626098;
CNPq grant 305974/2018-1; FAPERJ grants E-26/202.915/2019, E-26/211.134/2019.




\bibliographystyle{IEEEtran}
\bibliography{paper}
%
%
%

%

\vspace{-1.25cm}
\begin{IEEEbiographynophoto}{Shaoyu Chen}
is a Ph.D. candidate in the Computer Science and Engineering Dept. at NYU. He received the B.Eng. degree in computer science from HKUST. His research focuses on urban data visualization and virtual reality.
\end{IEEEbiographynophoto}

\vspace{-1.3cm}
\begin{IEEEbiographynophoto}{Fabio Miranda}
is an Assistant Professor in the Computer Science Dept. at UIC. He received the Ph.D. degree in computer science from NYU and the M.Sc. degree in computer science from PUC-Rio. His research focuses on large scale data analysis, data structures, and urban data visualization.
\end{IEEEbiographynophoto}
\vspace{-1.25cm}

\begin{IEEEbiographynophoto}{Nivan Ferreira}
is an Assistant Professor at UFPE in Brazil. He received a BSc in Computer Science and MSc in Mathematics from UFPE and PhD in Computer Science from NYU. Nivan was also a Post-Doc at the Department of Computer Science at the University of Arizona. Nivan's research focuses on many aspects of interactive data visualization, in particular systems and techniques for analysis spatiotemporal datasets.
\end{IEEEbiographynophoto}
\vspace{-1.25cm}

\begin{IEEEbiographynophoto}{Marcos Lage}
is a professor in the Dept. of Computer Science at UFF in Brazil, and is one of the principal investigators  of  the  Prograf  lab.  His  research interests  include  aspects  of  visual  computing,especially scientific and information visualization,numerical simulations, geometry processing, and topological data structures. He has a Ph.D. in applied mathematics from PUC-Rio.
\end{IEEEbiographynophoto}
\vspace{-1.25cm}

\begin{IEEEbiographynophoto}{Harish Doraiswamy}
is a Research Scientist at the NYU Center for Data Science. He received his Ph.D. in Computer Science and Engineering from the Indian Institute of Science, Bangalore. His research interests lie in the intersection of computational topology, visualization, and data management. His recent research focuses on the analyses of large spatio-temporal datasets from urban environments.
\end{IEEEbiographynophoto}
\vspace{-1.25cm}

\begin{IEEEbiographynophoto}{Corinne Brenner}
is a Ph.D. student in Educational Communication and Technology at NYU, studying educational applications of immersive media. She holds a B.A. in Psychology from Cornell University and an M.Sc. in Social Psychology from the University of Amsterdam
\end{IEEEbiographynophoto}
\vspace{-1.25cm}

\begin{IEEEbiographynophoto}{Connor Defanti}
is a technical consultant for Numerati Partners LLC. He received his Ph.D. at NYU's department of Mathematics and Computer Science and his B.S. at Caltech. His research focus has been multi-person, immersive virtual reality systems.
\end{IEEEbiographynophoto}
\vspace{-1.25cm}

\begin{IEEEbiographynophoto}{Michael Koutsoubis}
is a VR/MR Specialist at Kohn Pedersen Fox. With a background in architectural design, he focuses his professional career on utilizing emerging technology to improve how businesses communicate spatial ideas. He is also a Co-founder of Mythic VR.
\end{IEEEbiographynophoto}
\vspace{-1.25cm}

\begin{IEEEbiographynophoto}{Luc Wilson}
is  a  senior  associate  principal  at Kohn  Pedersen  Fox  and  the  director  of  KPF Urban Interface, a think-tank focused on urban analytics.  He is also an Adjunct Assistant Professor at Columbia GSAPP, and an Adjunct Course Advisor in the Center for Data Science at NYU. He earned his M.Arch. from Columbia University.
\end{IEEEbiographynophoto}
\vspace{-1.25cm}

\begin{IEEEbiographynophoto}{Ken Perlin}
is a professor in the Department of Computer Science at NYU. His research interests include future reality, graphics and animation, user interfaces and education. He received an Academy Award for Technical Achievement from the Academy of Motion Picture Arts and Sciences for his noise and turbulence procedural texturing techniques, which are widely used in feature films and television. 
\end{IEEEbiographynophoto}
\vspace{-1.25cm}

\begin{IEEEbiographynophoto}{Claudio Silva}
is a professor of computer science and engineering and data science with NYU. His research lies in the intersection of visualization, data analysis, and geometric computing, and recently has focused on urban and sports data. He has received a number of awards: IEEE Fellow, IEEE Visualization Technical Achievement Award, and elected chair of the IEEE Visualization \& Graphics Technical Committee.
\end{IEEEbiographynophoto}





\end{document}